\documentclass[a4paper,11pt]{article}
\pdfoutput=1 

\usepackage{jcappub} 

\usepackage[T1]{fontenc} 

\usepackage{graphicx}
\usepackage{color}
\usepackage[dvipsnames]{xcolor}
\usepackage{xspace}
\usepackage{multirow}
\usepackage{cleveref}
\usepackage[normalem]{ulem}
\usepackage{booktabs}
\usepackage{array}
\usepackage[all]{nowidow}

\usepackage{placeins}

\DeclareMathAlphabet\mathbfcal{OMS}{cmsy}{b}{n}

\newcommand{\fulleqref}[1]{Equation~\eqref{#1}}

\newcommand{\tquote}[1]{``#1''}

\newcommand\ion[2]{#1$\,${\scshape{#2}}}

\providecommand{\lcdm}{\ensuremath{\Lambda\mathrm{CDM}}\xspace}
\providecommand{\lcdmnu}{\ensuremath{\Lambda\mathrm{CDM}\nu}\xspace}
\newcommand{\lya}{\ensuremath{\text{Lyman-}\texorpdfstring{\alpha}{alpha}}\xspace}

\newcommand{\citepy}{PY15~\cite{Palanque-Delabrouille:2015pga}\xspace}
\newcommand{\citebp}{BP14~\cite{Borde2014}\xspace}

\makeatletter

\newcommand*{\@rowstyle}{}

\newcommand*{\rowstyle}[1]{
  \gdef\@rowstyle{#1}%
  \@rowstyle\ignorespaces%
}

\newcolumntype{=}{
  >{\gdef\@rowstyle{}}%
}

\newcolumntype{+}{
  >{\@rowstyle}%
}

\title{\boldmath Hints, neutrino bounds, and WDM constraints from SDSS DR14 Lyman-$\alpha$ and Planck full-survey data}

\author[a]{Nathalie Palanque-Delabrouille,}
\author[a]{Christophe Y\`eche,}
\author[b]{Nils Sch\"oneberg,}
\author[b]{Julien Lesgourgues,}
\author[a]{Michael Walther,}
\author[a]{Sol\`ene Chabanier,}
\author[a]{Eric Armengaud}

\affiliation[a]{\textit{IRFU, CEA}, Universit\'e Paris-Saclay, \\ F-91191 Gif-sur-Yvette, France}
\affiliation[b]{RWTH Aachen University, \textit{Institute for Theoretical Particle Physics and Cosmology (TTK)},  \\Sommerfeldstrasse 16, 52074, Aachen, Germany}

\emailAdd{nathalie.palanque-delabrouille@cea.fr}
\emailAdd{christophe.yeche@cea.fr}
\emailAdd{schoeneberg@physik.rwth-aachen.de}
\emailAdd{lesgourg@physik.rwth-aachen.de}
\emailAdd{michael.walther@cea.fr}
\emailAdd{solene.chabanier@cea.fr}
\emailAdd{eric.armengaud@cea.fr}

\abstract{The \lya forest 1D flux power spectrum is a powerful probe of several cosmological parameters. 
Assuming a \lcdm cosmology including massive neutrinos,  we find that the latest  SDSS DR14  BOSS and eBOSS \lya forest data is in very good agreement with current weak lensing constraints on $(\Omega_m, \sigma_8)$ and has the same small level of tension with Planck. We did not identify a systematic effect in the data analysis that could explain this small tension, but we show that it can be reduced in extended cosmological models where the spectral index is not the same on the very different times and scales probed by CMB and \lya data.
A particular case is that of a \lcdm model including a running of the spectral index on top of massive neutrinos. With combined \lya and Planck data, we find a slight (3$\sigma$) preference for negative running, $\alpha_s= -0.010 \pm 0.004$ (68\%CL). Neutrino mass bounds are found to be robust against different assumptions. In the \lcdm model with running, we find $\sum m_\nu <0.11$~eV at the 95\% confidence level for combined \lya and Planck (temperature and polarisation) data, or  $\sum m_\nu < 0.09$~eV when adding CMB lensing and BAO data. We further provide strong and nearly model-independent bounds on the mass of thermal warm dark matter.  For a conservative configuration consisting of  SDSS data restricted to $z<4.5$ combined with XQ-100 \lya data, we find $m_X > 5.3\;\mathrm{keV}$ (95\%CL).}

\begin{document}
\hfill{\small TTK-19-48}\newline
\maketitle
\flushbottom

\setlength{\parskip}{0.5\baselineskip}

\section{Introduction} 
\label{sec:intro}

Modern cosmological experiments have delivered an impressive amount of high-quality data on several observables, related either to cosmic microwave background (CMB) anisotropies, to the Large Scale Structure of the Universe, or to its expansion history. In many cases, the measurements are so precise that the interpretation of the data is limited by systematic rather than statistical errors. Nevertheless, a consistent picture is emerging, with the minimal 6-parameter \lcdm model standing out as the simplest explanation for most observations. 

There are still very strong reasons for gathering more and better data. First, the current status of the \lcdm model is not entirely clear, with intriguing and persistent hints of anomalies like the small-scale dark matter crisis, the Hubble tension, and the $\sigma_8$ tension. To better understand what is going on, we need to combine several types of observations, sensitive to different systematics. The comparison of different and independent probes is the best way to get a clue on the origin of these tensions. Second, on the basis of laboratory experiments and theoretical modelling, we believe that the total neutrino mass very likely impacts cosmological observables at a level which is not very far from the current detection threshold. Third, the possible detection of any signature beyond the simplest paradigm of a cosmological constant, of some plain cold dark matter, of slow-roll inflation and Einstein's gravity would have a very deep impact on our quest for new physics. 

The \lya forest flux power spectrum is a powerful tool to study clustering in the universe at redshifts 2 to 6, on scales that are strongly non-linear today, but were only mildly non-linear at such high redshifts. Since it extends the lever arm of other cosmological probes towards smaller scales, it has often been used to constraint the parameters of models affecting only the smallest scales -- like warm dark matter (WDM) \cite{Narayanan:2000tp,Viel:2005qj,Seljak:2006qw,Boyarsky:2008xj} or interacting dark matter \cite{Dvorkin:2013cea,Xu:2018efh,Garny:2018byk} -- or altering the global shape of the power spectrum -- like non-slow-roll inflation or massive neutrino free-streaming \cite{Seljak:2004xh,Seljak:2006bg}. Recent data from the \lya flux power spectrum provide some of the strongest existing bounds on the mass of WDM \cite{Baur2016,Baur2017,Irsic2017,Irsic2017b,Armengaud2017}. In combination with CMB data, they also provide the strongest cosmological neutrino mass bounds to date \cite{Seljak2006,viel2010,Palanque-Delabrouille:2015pga,Palanque-Delabrouille:2014jca,Yeche2017}.

A new high precision measurement of the one-dimensional \lya flux power spectrum was recently published~\cite{Chabanier2019}, using  43\,751 high-quality quasar spectra from the Data Release 14 of the BOSS and eBOSS collaborations~\cite{Dawson2013,Dawson2016}. It covers thirteen redshift bins from $z= 2.2$ to 4.6, up to the wavenumber $k = 0.02\; \text{s\;km}^{-1}$. This work presents a first cosmological interpretation of this  data set. For this purpose, we have assembled a grid of 138 hydrodynamical simulations with a resolution equivalent to $3 \times 3072^3$ particles in a $(100\text{Mpc/h})^3$ box,  which we obtain using a splicing technique, extending the grid from \cite{Borde2014} used in previous work. We have built a likelihood for the new data set, and two analysis pipelines for cosmological parameter extraction.

In section \ref{sec:data} we describe the data sets that we employ in this work. In section \ref{sec:lyasims} we explain our settings and assumptions for numerical simulations and for the astrophysical modeling of the intergalactic medium (IGM). In section \ref{sec:freq_vs_bay} we have a short discussion on the frequentist and Bayesian interpretation methodologies. Our results are presented in section \ref{sec:results}. We start from the analysis of \lya data alone in section~\ref{sec:LyaCMB}, and we discuss its compatibility with Planck and BAO data in section \ref{sec:tension}. In section \ref{sec:lyacmb} we combine the \lya, CMB, and BAO data for the \lcdm model with massive neutrinos, and we show that the compatibility between \lya and Planck data could be improved in cosmological models where the effective slope of the spectrum is different on CMB and \lya scales. In section \ref{sec:running} we present our bounds for an example of such a model, involving a running of the spectral index. We present our neutrino (resp. WDM) mass bounds and discuss  their robustness in section \ref{sec:mnu} (resp. \ref{sec:wdm}). Our conclusion are summarized in \ref{sec:conclusion}.

\section{Data and Methodology}
In this section, we first present the data sets we use. We then recall the modeling of the \lya forest developed in earlier works and introduce several new improvements. We finally briefly describe the two interpretation methodologies used throughout this work, which allow us to test the robustness of the obtained results. 

\subsection{Data} \label{sec:data}

\subsubsection{\lya forest 1D flux power spectrum } \label{sec:data_lya}

\begin{figure}[t]
\begin{center}
\epsfig{figure= 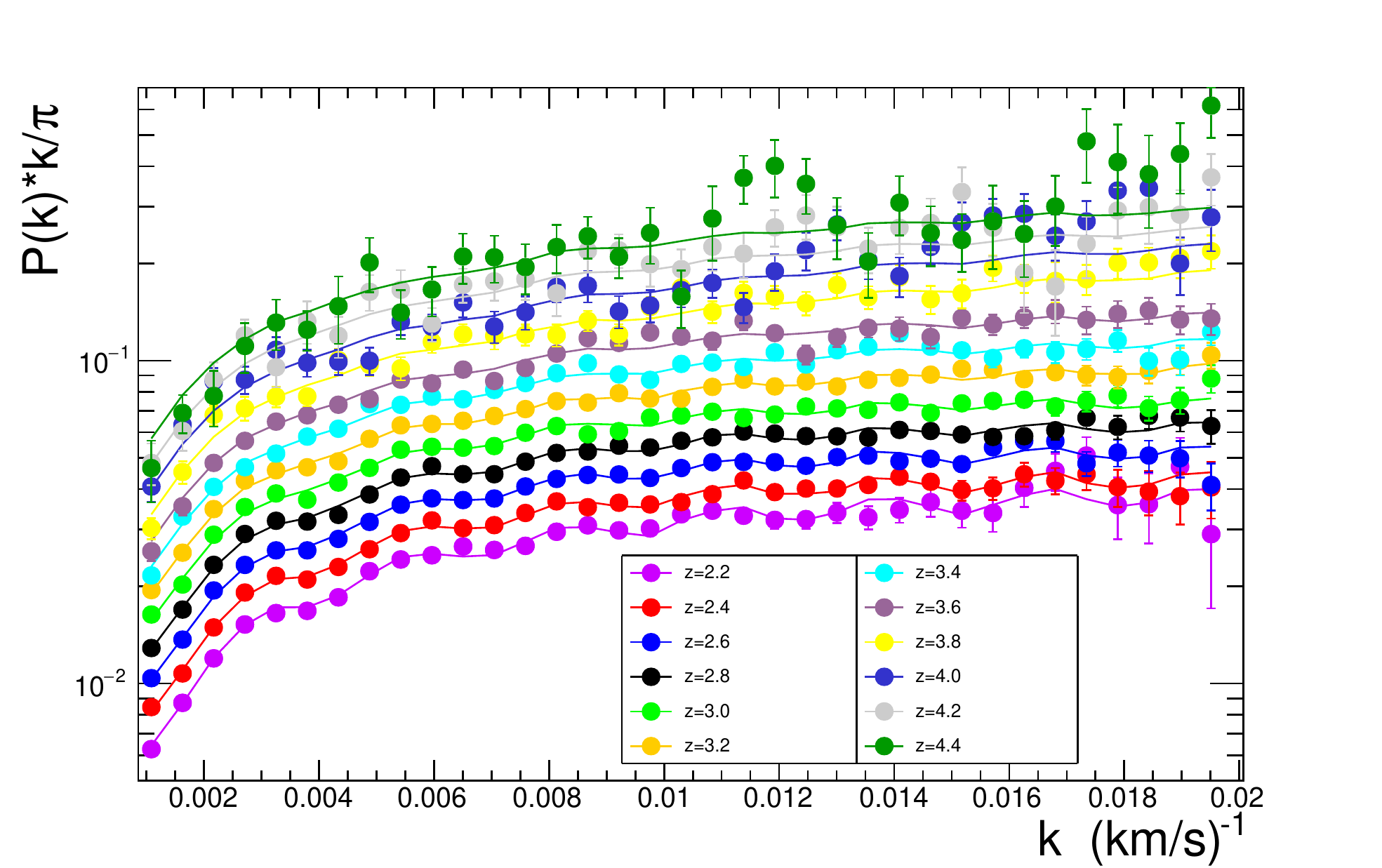,width = .9\textwidth}\\
\epsfig{figure= 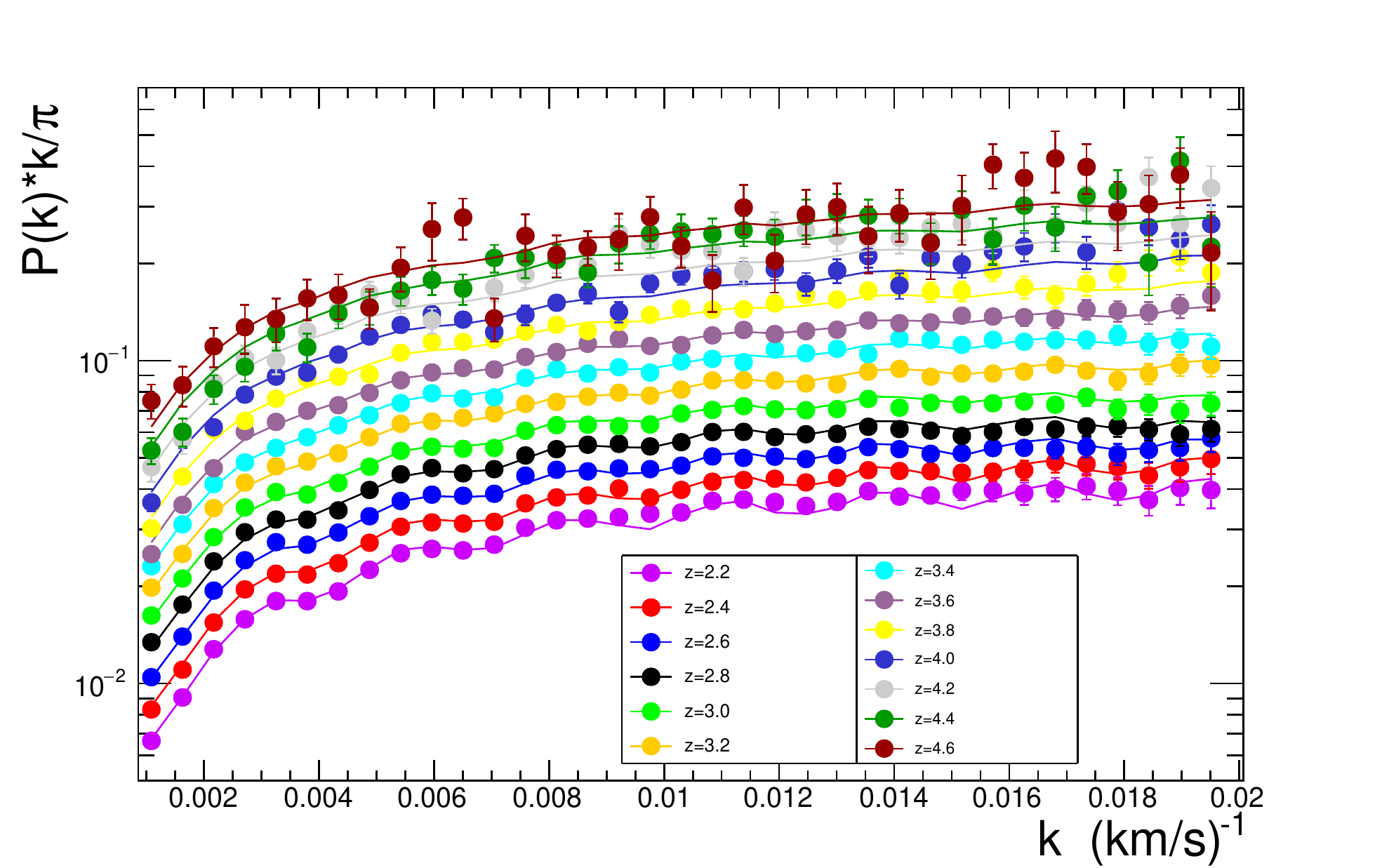,width = .9\textwidth}\\
\caption{\label{fig:Lyafit} \textbf{Top:} BOSS DR9 \lya data from~\cite{Palanque-Delabrouille:2015pga} and best-fit $\Lambda$CDM + $\sum m_\nu$ model. \textbf{Bottom:} eBOSS DR14 \lya data from~\cite{Chabanier2019} and best-fit $\Lambda$CDM + $\sum m_\nu$ model. }
\end{center}
\end{figure}

We use the 1D \lya flux power spectrum measurement from~\citepy{}, which is based on the DR14 BOSS and eBOSS data of the Sloan Digital Sky Survey (SDSS). The data consist of a subsample of $43\,751$ quasars selected from a parent sample of $180\,413$ quasars~\cite{Paris2018,Ross2012,Myers2015,Palanque2016} according to the spectrum quality. This sample  improves over the one  from SDSS BOSS DR9 used  in previous works both in statistical precision (achieving a reduction by a factor of two) and in redshift coverage. We now have 13 equally-spaced redshift bins covering the range from $z=2.2$ to $4.6$,  as opposed to 12 bins from $z=2.2$ to $4.4$ in \cite{Palanque-Delabrouille2015}. These data also come with systematic uncertainties that result from an in-depth study of relevant observational (resolution, noise, sky subtraction, continuum fitting) as well as astrophysical effects (damped \lya systems, broad absorption lines, metal contamination). We thus perform the analysis on 435 \lya data points, spread evenly over 35 bins in $k$ space (from $k=0.0011\;{\rm s\;km^{-1}}$ to  $0.0195\; {\rm s\;km^{-1}}$) and 13 bins in redshift space. Figure~\ref{fig:Lyafit} illustrates the SDSS DR14 \lya 1D flux power spectrum measurement  and the best-fit $\Lambda$CDM + $\sum m_\nu$ model of section~\ref{sec:LyaCMB}.

We also explore an extended \lya data set by considering the 1D \lya flux power spectrum measurement from the one hundred XQ-100 quasars~\cite{Yeche2017}. These data cover three redshift bins at $z=3.2$, 3.6, and 3.9. They  exhibit a better resolution than the SDSS data, and thus allow an extension of the analysis to slightly higher $k$ modes, namely to $k=0.05\;{\rm s\;km^{-1}}$ for $z=3.2$, $k=0.06\;{\rm s\;km^{-1}}$ for $z=3.6$, and $k = 0.07 \;{\rm s\;km^{-1}}$ for $z=3.9$.

\subsubsection{Cosmic microwave background} \label{sec:data_cmb}

For the CMB data, we use the Planck 2018 data sets. The likelihoods for the Bayesian approach are described in \cite{Aghanim:2019ame}, and the corresponding publicly available chains are used for the frequentist approach. We  always use the full low-$\ell$  and high-$\ell$ temperature and polarization data, denoted as \tquote{P18}. When specified, we also include the Planck 2018 CMB lensing data, denoted as  \tquote{lens.}. 

\subsubsection{Baryon acoustic oscillations} \label{sec:data_bao}

We also combine CMB data with measurements of the BAO scale by 6dFGS~\cite{Beutler2011}, SDSS main galaxy sample~\cite{Ross:2013vla}, BOSS-LOWZ, and CMASS from DR12~\cite{Alam2017}. Theses measurements are henceforth globally denoted \tquote{BAO}. The additional constraints that these measurement provide on cosmological parameters are included in the present work with their full correlation with CMB data. For the frequentist approach described in section~\ref{method:freq}, both CMB and BAO constraints are taken from the Markov chains publicly available through the official Planck legacy archive. For the Bayesian approach, the constraints are directly derived from the likelihoods available through {\sc MontePython 3}.

\subsection{Modeling of the \texorpdfstring{\lya}{Lyman-Alpha} forest 1D flux power spectrum}

\label{sec:lyasims}

\begin{table}[t]
    \begin{center}
    \begin{tabular}{p{2.9cm}ll}
    \hline
    \hline
    Parameter& Definition \\
    \hline
    $\Omega_m$ \dotfill & Matter fraction today (compared to critical density) \\
    $H_0$ \dotfill &  Expansion rate today in km s$^{-1}$ Mpc$^{-1}$ \\
    $\sigma_8$ \dotfill & RMS matter fluctuation amplitude today in linear theory \\
    $n_s$ \dotfill & Scalar spectral index \\
    $\sum m_\nu$ \dotfill & Sum of neutrino masses \\
    $m_X$ \dotfill & Mass of thermal relic WDM particle \\
    $\alpha_s$  \dotfill & Running of the power spectrum scalar index \\
    \hline
    $T_0(z=3)$ \dotfill & Normalization temperature of IGM at $z=3$ \\
    $\gamma(z=3)$ \dotfill &  Logarithmic slope of  $\delta$-dependence of IGM temperature at $z=3$ \\
    $\eta^{T}(z<3)$ \dotfill &  Logarithmic slope of  $z$-dependence of $T_0$ for $z<3$\\
    $\eta^{T}(z>3)$ \dotfill &  Logarithmic slope of  $z$-dependence of $T_0$ for $z>3$\\
    $\eta^{\gamma}$  \dotfill &  Logarithmic slope of  $z$-dependence of $\gamma$ \\
    $A^\tau$  \dotfill &  Amplitude of the effective optical depth of Ly$\alpha$ absorption $\tau_{\rm eff}$ \\
    $\eta^\tau$  \dotfill &  Logarithmic slope of  redshift dependence of $\tau_{\rm eff}$   \\
    $f_ {\rm{Si\,III}}$  \dotfill &  Fraction of  \ion{Si}{III} absorption relative to Ly$\alpha$ absorption\\
    $f_ {\rm{Si\,II}}$  \dotfill &  Fraction of  \ion{Si}{II} absorption relative to Ly$\alpha$ absorption \\
    $z_{\rm reio}$ \dotfill & Redshift of reionization\\
    \hline
    $A^\mathrm{splice}$ \dotfill & Amplitude of splicing correction \\
    $\eta^\mathrm{splice}$ \dotfill & Small-scale slope  of splicing correction\\
    $A^\mathrm{SN}$ \dotfill & Amplitude of supernova feedback correction\\
    $A^\mathrm{AGN}$ \dotfill & Amplitude of AGN feedback correction \\
    $A^\mathrm{UVfluct}$ \dotfill & Amplitude of UV fluctuation correction\\
    $A^{n,i}$ \dotfill & Amplitude of  noise power correction for redshift bin $i$\\
    \hline
    \end{tabular}
    \end{center}
    \caption{Definition of the parameters used throughout this work, sorted in three categories: cosmological, astrophysical, and nuisance. The latter two sets of parameters describe  corrections that are only applied to the \lya flux power spectrum.}
    \label{tab:astroparam}
\end{table}

To predict the theoretical \lya flux power spectrum, we use the set of simulations extensively described in \citebp{}  for the initial grid, in \cite{Rossi2014} for the active neutrino sector and in \citepy{} for systematic studies related to these simulations. The WDM extension of the grid was introduced in~\cite{Baur2016}. We present here a brief summary. The simulations are run using a  parallel tree smoothed particle hydrodynamics (tree-SPH) code {\sc Gadget-3}, an updated version of the public code {\sc Gadget-2} \cite{Springel2001,Springel2005}. The simulations are started at $z=30$, with initial transfer functions and power spectra computed with {\sc CAMB}~\citep{Lewis2000}, and initial particle displacements generated with second-order Lagrangian perturbation theory {\sc 2lpt}\footnote{\url{http://cosmo.nyu.edu/roman/2LPT/}}. We include three particle types: collisionless dark matter, gas and, when relevant, mass-degenerate neutrinos. 
We showed in \cite{Palanque-Delabrouille2015} that considering inverted or normal neutrino mass hierarchy yields a flux power spectrum that differs by less than 0.05\%  from the degenerate-mass scenario, a level ten times below the simulation statistical uncertainties and almost two orders of magnitude below the data uncertainties. The degenerate-mass hypothesis is thus highly justified. We use the quick-Ly$\alpha$ option to convert gas particles with overdensities exceeding $10^3$ and temperature below $10^5$~K into stars. The simulations cover the volume of a periodic  100~Mpc$/h$ box, containing the equivalent of $3072^3$ particles of each type. Following a method originally suggested in \cite{McDonald2003}, we obtain this resolution by splicing together large-volume  and  high-resolution simulations, using a transition simulation that corrects the large-box simulation for its lack of coupling between small and large modes, and the high-resolution simulation for its small volume. We studied the accuracy of the splicing technique in \cite{Palanque-Delabrouille2015} and \cite{Palanque-Delabrouille:2015pga}, and we correct for residual biases by the  nuisance parameters $A^\mathrm{splice}$ and $\eta^\mathrm{splice}$. Snapshots are produced at regular intervals in redshift from $z = 4.6$ to 2.2, with $\Delta z = 0.2$, thus corresponding to the same redshift bins as for the \lya data.

The dependence of the \lya flux power spectrum on the  parameters of interest is modeled by a Taylor expansion within a set of  cosmological  $\{\sigma_8, n_s, H_0, \Omega_m, \sum m_\nu$ or   $1/m_X, \alpha_s \}$ and  effective astrophysical  $\{T_0(z=3),\gamma(z=3),A^\tau,\eta^\tau\}$ parameters. The grid of simulations required for this interpolation consists of a reference simulation  centered on the Planck 2013 best-fit cosmology \cite{Ade:2013zuv} and an IGM thermal history in agreement with \cite{Meiksin2009, Becker2011}, completed by simulations where one or two parameters   at a time are given off-centered values. 
These simulations are used to compute a full second-order Taylor expansion around the   \lya flux power spectrum measured for the reference case. 

The cosmological parameters cover the range $H_0=67.5\pm 5~{\rm km\,s^{-1}\,Mpc^{-1}}$, $\Omega_M=0.31\pm0.05$, $n_s=0.96\pm0.05$,  $\sigma_8=0.83\pm0.05$. In all the runs, we keep $\Omega_b = 0.0221$. To constrain neutrino masses, additional simulations are run with $\sum m_\nu=0.4$ or $0.8$~eV.  Where WDM is assumed, the dark matter particles are thermal relics with masses $m_X=2.5$ or $5.0$~keV. Since \lcdm is reproduced for $m_X \to \infty$, we use  $1/m_X$ instead of $m_X$ in the Taylor expansion, and the simulations therefore probe $1\,{\rm keV}/m_X = 0$, $0.2$ and $0.4$. 
When running of the scalar index is assumed, we use a set of simulations where the running parameter $\alpha_s\equiv dn_s/d\ln k$ is fixed to $\pm 0.04$. This allows us to introduce running directly in the Taylor expansion and thus fully account for its impact  on the 1D \lya flux power spectrum, whether direct or through cross-correlations with the other parameters. This feature was not included in the previous analysis of~\citepy{}, where running was included a posteriori  though a change of $n_s$ between large (CMB) and small (\lya) scales according to the relation $n_s(k) = n_s(k_p) + \alpha_s \times \ln (k/k_p)$.

Astrophysical parameters are varied in the simulations by modifying the UV background. The photoheating rates are varied to cover the range  $\gamma(z=3)=1.3\pm 0.3$ and $T_0 (z=3) = 14000\pm 7000\,$K in the outputs, where the IGM temperature is modeled for each redshift according to the usual relation $T = T_0\times (\rho/\langle\rho\rangle)^{\gamma-1}$ \cite{Hui:1997dp}. The photo-ionization rate of each simulation was fixed by requiring the effective optical depth at each redshift to follow the empirical law $\tau_{\rm eff}(z) =  A^\tau  (1+z) ^{\eta^\tau}$, with $A^\tau=0.0025 \pm 0.0020$ and $\eta^\tau=3.7\pm0.4$~in agreement with~\citep{Meiksin2009}, and where $\tau_{\rm eff}(z)$ is defined with respect to the mean flux as $\tau_\mathrm{eff}(z) \equiv -\ln \langle F(z) \rangle$. This renormalization was done at the post-processing stage, as justified in \cite{Theuns2005}, allowing us to model  the impact of different scalings without running new simulations. 

The full modeling of the \lya flux power spectrum includes further parameters to allow for additional freedom in the IGM thermal history and to account for remaining uncertainties or biases. We thus allow $T_0(z)$ and $\gamma(z)$ to vary as $(1+z)^\eta$, with a broken power law at $z=3$ for $T_0$ and a single power law for $\gamma$, as done in several previous works and in agreement with the recent works of \citep{Walther2019,Puchwein:2014zsa}. We also include two amplitudes for the correlated absorptions by \lya with \ion{Si}{II} and \ion{Si}{III}, which we use as multiplicative corrections to the flux power spectrum.  The study of WDM requires one to be able to lift the degeneracy between Jeans smoothing and WDM free streaming. Therefore, for WDM studies, we account for changes in the redshift of reionization  by adding a nuisance term that reproduces the impact of $z_{\rm reio}$ on the redshift and mode dependence of the flux power spectrum, as was done in~\cite{Baur2016}.  The remaining parameters are nuisance parameters that allow us to account for uncertainties or corrections related to noise in the data, spectrograph resolution,  bias from the splicing technique,  UV fluctuations in the IGM, residual contamination from unmasked DLA, and supernova and AGN feedbacks. Details on the fit parameters and on the dependence with scale and redshift of the nuisance parameters can be found  in \citepy{}.  In this work we introduce an improved modeling of the AGN feedbacks derived from the detailed study of \cite{Chabanier2019b}. The authors measure a significant and well understood impact of AGN feedback on the \lya forest properties, which suppresses the \lya 1D flux power spectrum by up to 6\% on large scales, and of order 1\% on the smallest scales probed by the eBOSS data. The SN feedback acts on similar scales, and partially compensates the effect of AGNs. We  adopt the new more accurate model from~\cite{Chabanier2019b} to correct the predicted power spectrum for the AGN feedback and we use the same study as before from~\cite{viel12} to correct for the SN feedback. For both feedbacks, we apply a Gaussian prior around the central value of the correction (see table~\ref{tab:prior}). 
The definition of all parameters that describe the \lya flux power spectrum can be found in table \ref{tab:astroparam}.

\subsection{Interpretation methodology}\label{sec:freq_vs_bay}
To assess the robustness of the results, we use two different interpretation methodologies; frequentist and Bayesian. We briefly explain below how we combine the \lya and CMB studies in both cases.

\subsubsection{Frequentist interpretation} \label{method:freq}

Our determination of the coverage intervals of unknown cosmological parameters is based on the \tquote{classical} confidence level method originally defined by \cite{Neyman1937}. We start with the likelihood ${\cal L}\bigl(x,\sigma_x;\Theta)$, for a given cosmological model defined by the $n$ cosmological, astrophysical and nuisance parameters $\Theta=(\theta_{1},\ldots,\theta_{n})$, and for data measurements $x$ with Gaussian experimental errors $\sigma_{x}$.  In the rest of this paper, we adopt a $\chi^2$ notation, which means that the following quantity is minimized:
\begin{equation}
\chi^2(x,\sigma_x;\Theta) = -2 \ln ({\cal L}(x,\sigma_x;\Theta))~.
\label{eq:chi2}
\end{equation}
We first determine the minimum $\chi^2_\mathrm{min}$ of $\chi^2(x,\sigma_{x};\Theta)$ leaving  all the cosmological parameters free. Then, to set a confidence level (CL) on any individual cosmological parameter $\theta_i$, we scan the variable $\theta_i$: for each fixed value of $\theta_i$, we minimize again $\chi^2(x,\sigma_{x};\Theta)$ but with $n-1$ free parameters. The $\chi^2$ difference, $\Delta \chi^2(\theta_i)$, between the new minimum and  $\chi^2_\mathrm{min}$, allows us to compute the CL on the variable, assuming that the experimental errors are Gaussian,
\begin{equation}
{\rm CL}(\theta_i) = 1-\int_{\Delta \chi^2(\theta_i)}^{\infty}  f_{\chi^2}(t;N_{dof}) dt~,
\label{Eq:CL}
\end{equation}
with the $\chi^2$ distribution 
\begin{equation}
 f_{\chi^2}(t;N_{dof})=\frac{e^{-t/2}t^{N_{dof}/2 -  1}}{\sqrt{2^{N_{dof}}} \Gamma(N_{dof}/2)}   \label{Eq:chi2}~,
\end{equation}
where $\Gamma$ is the Gamma function and the number of degrees of freedom $N_{dof}$
is equal to 1.
This profiling method can be easily extended to two variables. In this case, the minimizations are
performed for $n-2$ free parameters and the confidence level ${\rm CL}(\theta_i,\theta_j)$ is derived from \fulleqref{Eq:CL} with $N_{dof}=2$.

In this paper we also combine the  $\chi^2$ derived from the Ly$\alpha$ likelihood with that of Planck. In the frequentist analysis, we do not use the Planck likelihoods directly, but we use the Markov chains available in the official Planck\footnote{\tt https://wiki.cosmos.esa.int/planck-legacy-archive/index.php/Cosmological\_Parameters } repositories instead. 
 For instance, for the  Planck 2018 TT+TE+EE configuration with massive neutrino, we take the chains from the directory \texttt{base\_mnu/plikHM\_TTTEEE\_lowl\_lowE}, which we reduce to the cosmological parameters  $\{\sigma_8, n_s, \Omega_m, H_0, \sum m_\nu\}$ that are relevant for our analysis. The distribution of the chain elements allow us to estimate the posterior probability distributions for each parameter and  the correlations between  parameters.  The flat positive prior applied to  $\sum m_\nu$ causes a distortion of all posterior probability distributions, in particular for $\sum m_\nu$, and thus also for $\{\sigma_8, \Omega_m$\}, which are strongly correlated with $\sum m_\nu$.  The posterior probability distribution becomes asymmetric and cannot be modeled by a simple Gaussian distribution. To account for such effects, we first apply a Principal Component Analysis on the reduced chain that allows us to determine  the linearly uncorrelated variables, called the principal components. We then model the distribution of each principal component  by several asymmetric Gaussians. We validated this strategy for a few configurations by comparing the limits obtained on $\sum m_\nu$ with this modeling with the limits derived directly from the MCMC approach using the full likelihood. The agreement between the two approaches was typically at the level of a few percent.

\subsubsection{Bayesian interpretation}\label{sec:bay}

\begin{table}[t]
    \centering
    \begin{tabular}{c|c | c | c | c}
         Parameter & Minimum & Maximum & Gaussian Mean & Gaussian Sigma\\
         \toprule
         $f_\mathrm{SiIII}$ & -0.2 & 0.2 & - & -\\
         $f_\mathrm{SiII}$ & -2.0 & 2.0 & - & -\\
         $A^\tau$ & 0 & 1.5 & - & -\\
         $\eta^\tau$ & 0 & 7.0 & - & -\\
         $T_0(z=3)$ & 0 & 25000 & - & -\\
         $\gamma(z=3)$ & 0.3 & 2.0 & 1.3 & 0.3\\
         $\eta^{T}(z<3)$ & -5.0 & 2.0 & 1.0 & 2.0\\
         $\eta^{T}(z>3)$ & -10.0 & 2.0 & -2.0 & 3.0\\
         $\eta^{\gamma}$ & -5.0 & 2.0 & 0.1 & 1.0\\
         $z^\mathrm{reio}$ & 7.0 & 15.0 & - & -\\
         $A^\mathrm{splice}$ & -1.0 & 1.0 & 0.01 & 0.05\\
         $\eta^\mathrm{splice}$ & -40.0 & 40.0 & 0 & 2.5\\
         $A^\mathrm{SN}$ & 0.0 & 3.0 & 1.0 & 0.3\\
         $A^\mathrm{AGN}$ & 0.0 & 3.0 & 1.0 & 0.3\\
         $A^\mathrm{UVfluct}$ & 0.0 & 3.0 & 0 & 0.3\\
         $A^{n,i}$ & -2.0 & 2.0 & 0 & 0.02\\
    \end{tabular}
    \caption{Bayesian prior ranges on astrophysical/nuisance parameters. The priors are either flat (denoted with - signs), or Gaussian with given mean and standard deviation.  The $z_\mathrm{reio}$ parameter is only used for the thermal warm dark matter case.}
    \label{tab:prior}
\end{table}

The Bayesian determination of the credible intervals of the parameters is based on their full posterior distribution according to Bayes theorem. The posterior distribution represents the probability distribution of the estimation of the true parameter value, which was improved upon by the experiment compared to the prior estimation. To obtain the posterior distribution, we use the implementation of the Metropolis Hastings algorithm from \mbox{{\sc MontePython}~3~\cite{Brinckmann:2018cvx}} for our Monte-Carlo Markov Chains (MCMC). For each case, we run chains of an average total length of around 7 million steps, out of which on average around 1.5 million steps are accepted. We have explicitly checked that the convergence criteria of Gelman-Rubin  for any parameter of any run are at most $|R-1|<0.01$.

For our cosmological parameter basis we choose $\{\omega_b,\omega_\mathrm{cdm},100\theta_s,\ln 10^{10}A_s, n_s,\tau_\mathrm{reio}\}$, plus a total neutrino mass $\sum m_\nu$ split beteween three degenerate species. Since this corresponds to the parameter basis in the Planck analysis, one can directly see the improvement coming from the addition of \lya data. Many of the common degeneracies in CMB data analyses are removed by choosing appropriate flat priors (e.g. on $\theta_s$ instead~of~$H_0$). However, we checked explicitly (e.g. with runs assuming a flat prior on $H_0$ rather than $\theta_s$\,, $\sigma_8$ rather than $\ln 10^{10}A_s$, and $\Omega_m$ rather than $\omega_\mathrm{cdm}$) that the choice of the parameter basis does not significantly influence the results.

The prior range for each astrophysical or nuisance parameter is shown in table \ref{tab:prior}. Cosmological parameters are all allowed to vary freely without bounds (which amounts to assuming a flat prior in a wide range compared to the width of the posterior), except for two cases. We impose $\tau_\mathrm{reio}>0.004$, since smaller values imply a reionization happening below redshift $z=1$, which causes numerical problems and is now excluded by almost all data sets. Additionally, we impose the physical limitation  $\sum m_\nu > 0$. Note that oscillation experiments constrain $\sum m_\nu$ to be larger than at least $\sqrt{\lvert \Delta m_{32}^2\rvert} \sim 0.05$eV (see e.g. \cite{Tanabashi:2018oca}, section 14). Assuming a lower prior edge $\sum m_\nu > 0$ still makes sense: it amounts to studying limits on neutrino masses with minimal assumptions on neutrino decoupling and neutrino stability on cosmological time scales (which is not granted, e.g. \cite{Beacom:2004yd,Chacko:2019nej}). One could adopt a different point of view and derive  cosmological bounds under the assumption that neutrinos decouple in the standard way and are fully stable, such that the oscillation prior $\sum m_\nu > 0.05$eV applies. This would lead to slightly weaker cosmological bounds on $\sum m_\nu$ \cite{RoyChoudhury:2019hls}. Both approaches are consistent but simply address different questions. In this work, we only explore the first prior choice.

The 95\% CL for the sum of the neutrino masses is always derived as containing 95\% of the full posterior integral, starting from the lower bound of the posterior. The quoted $1\sigma$ uncertainties on the other parameters are those that contain $\sim 68.3\%$ of the posterior centered around the mean value. Usually the 1D posteriors are close enough to being Gaussian that we do not display  information other than the mean and sigma values.

\section{Results} \label{sec:results}

We present below the results obtained with both methodologies. We  first focus on the constraints from \lya data alone, and assess their compatibility with CMB constraints. In particular, we discuss a mild tension in the $\Omega_m$ parameter between the CMB and \lya data. We then combine the data sets, assuming either a minimal cosmology (\lcdm with massive neutrinos) or some basic extensions that may reduce this mild tension, with a particular focus on models with a running of the primordial spectral index. We  discuss active neutrino mass bounds and their robustness against different assumptions. Finally, we present the constraints obtained for warm dark matter in the form of a thermal relic of mass $m_X$ or in the form of a non-resonantly-produced sterile neutrino of mass $m_s$, which are robust with respect to the aforementioned tension.

\subsection{Cosmological constraints from \texorpdfstring{\lya}{Lyman-alpha} data alone}

\label{sec:LyaCMB}

\begin{table}[t]
\begin{center}
\begin{tabular}{lcccc}
\hline\hline
&\multicolumn{2}{c}{Frequentist} & \multicolumn{2}{c}{Bayesian} \\
& $m_\nu=0$  & Varying $m_\nu$  & $m_\nu=0$ & Varying $m_\nu$\\
 \hline \\[-10pt]

$T_0$ (z=3) {\scriptsize($10^3$K)}  & $8.5\pm1.9$   & $8.5\pm2.0$  & $8.6\pm1.9$  & $8.6\pm1.9$ \\[2pt]
$\gamma$                                        &  $0.93\pm0.14$   &  $0.93\pm0.14$     & $0.96\pm0.15$ & $0.98\pm0.14$\\ [2pt]
$\sigma_8$                                      &  $0.826\pm0.020$    & $0.826 \pm 0.021$     & $0.823\pm 0.022$ & $ 0.811\pm 0.024$\\[2pt]
$n_s$                                              &  $0.954\pm0.006$     &  $0.954\pm0.006$    &  $0.953 \pm 0.007$ & $0.955 \pm 0.007$\\[2pt]
$\Omega_m$  &  $0.269\pm0.009$     &  $0.269\pm0.009 $     &  $0.270\pm0.010$& $0.275 \pm 0.012$\\[2pt]
$\sum \! m_\nu$~{\scriptsize(eV , 95\% CL)} & - & $<0.58$ & - & $< 0.71$ \\[2pt]

\hline
\end{tabular}
\caption{\label{tab:results_lya_alone}Preferred astrophysical and cosmological parameter values (68.3\% confidence level) for the \lcdm + $m_\nu$ model, for \lya data combined with a Gaussian prior $H_0=67.3\pm1.0\,{\rm km\,s^{-1}\,Mpc^{-1}}$. } 
\end{center}
\end{table}

As was noted in \citepy{}, the \lya forest flux power spectrum only weakly depends on the Hubble parameter, and is unable to constrain $H_0$ by itself. When using \lya data alone, we thus adopt the same Gaussian prior constraint as in \cite{Palanque-Delabrouille2015, Baur2016, Baur2017, Chabanier2019}, which is taken from the Planck 2015 TT+lowP results~\cite{Planck2015}, namely $H_0 = 67.3 \pm 1.0\,{\rm km\;s^{-1}\;Mpc^{-1}}$. Note that most recent result from~\cite{Planck2018} using TT, TE, EE, low E + lensing gives $H_0 = 67.36 \pm 0.54$, in perfect agreement with the prior mentioned above.  In~\cite{Palanque-Delabrouille2015}, we explicitly checked that bounds on other parameters depend very weakly on the choice of $H_0$ prior. In particular, the \lya posteriors are \textit{not} significantly different when we combined the \lya data with an $H_0$ prior taken from the SH${}_0$ES results \cite{Riess:2019cxk}.

The IGM thermal history is one of the main sources of nuisance in this study. To encompass a large range of possible histories, we marginalize over the thermal parameters of table~\ref{tab:astroparam} to derive constrains either on $\sum m_\nu$ (sections~\ref{sec:LyaCMB} to \ref{sec:mnu}) or on the mass of a WDM thermal relic (section~\ref{sec:wdm}). We show in appendix~\ref{sec:appendixA} the recovered best-fit thermal history in each case. Because of the freedom allowed in the modeling, the uncertainties on the thermal parameters are large, and the $2\sigma$ range on $T_0$, $\gamma$ and mean flux overlaps with observational measurements.  To further test the robustness of our result, we also investigate how the constraints we derive depend on our hypotheses  and we investigate the impact of a different thermal model. As shown in appendix~\ref{sec:appendixA}, imposing a thermal model in agreement with a specific set of measurements~\cite{Becker2011} has little impact on the measured bounds.

As a sanity check on the interpretation methodologies, we  assess the compatibility between the Bayesian and frequentist best-fit values of the cosmological parameters for  \lya data alone, in the standard $\Lambda$CDM cosmological model with massless or massive neutrinos. As shown in table~\ref{tab:results_lya_alone}, the results for the two methodologies are in excellent agreement, both for the central value of the parameters and for their uncertainty. The flux power spectrum of the best-fit model is shown in figure~\ref{fig:Lyafit}. Compared to the previous work of \citepy{}, there is a noticeable improvement in the agreement between the data and the best-fit model, in particular for high~$k$ and high~$z$.  

 The best fit  on \lya data alone slightly differs from the one obtained on the DR9 data for two cosmological parameters: $n_s$, which increased from $0.938 \pm 0.010$ in \citepy{} to $0.954 \pm 0.006$, and $\Omega_m$, which decreased from $0.293\pm0.014$ to $0.269\pm0.009$.  We investigated the origin of this $\sim 1.5\,\sigma$ shift. Restricting the eBOSS data to the forests in common with those from DR9, we measure $n_s = 0.945 \pm 0.008$ and $\Omega_m = 0.278 \pm 0.015$. To further mimic the DR9 selection, we  then replace the automated catalogs of Broad Absorption Line quasars and Damped Lyman-$\alpha$ systems by the visual catalogs that were used in DR9. The fit on the resulting sample gives $n_s = 0.935 \pm 0.008$ and $\Omega_m = 0.282 \pm 0.015$, in good agreement with the cosmology obtained with the analysis of \citepy{} on the DR9 sample. This result concurs  with what was found in a similar study led in~\cite{Chabanier2019}. The change in the data is therefore at the origin of the small shift in best-fit cosmological parameters. 

\FloatBarrier
\subsection{Mild tension between Planck and \texorpdfstring{\lya}{Lyman-alpha} data \label{sec:tension}}

\begin{table}[t]
\begin{center}
\begin{tabular}{lcc}
\hline\hline
& P18 & P18  \\
&  &+ lens. +BAO  \\
 \hline \\[-10pt]

$\sigma_8$ &  $0.804\pm0.018$ &   $0.815 \pm 0.009$ \\[2pt]
$n_s$ &  $0.9630\pm0.0048$  &  $0.966 \pm 0.004$    \\[2pt]
$\Omega_m$  &  $0.321\pm0.014$   & $0.310 \pm 0.007$ \\[2pt]
$100\Omega_b$  &  $2.232\pm0.016$  & $2.241 \pm 0.014$   \\[2pt]
$H_0$~{\scriptsize(${\rm km~s^{-1}~Mpc^{-1}}$)}   &  $66.9 \pm 1.1$  & $67.81 \pm 0.5$    \\[2pt]
$\sum \! m_\nu$~{\scriptsize(eV , 95\% CL)} & $<0.286$ & $<0.113$ \\[2pt]

\hline
\end{tabular}
\caption{\label{tab:results_Planck} Preferred cosmological parameter values (68.3\% confidence level) fo the \lcdm + $\sum m_\nu$ model, fot Planck data alone or combined with BAO data, as extracted from Bayesian chains containing $\sim 6.5$ million points. The convergence criterion of the chains is \mbox{$|R-1| < 10^{-3}$} for all parameters, such that these chains are slightly more converged than the publicly available ones.}
\end{center}
\end{table}

\begin{figure}[t]
\begin{center}
\epsfig{figure=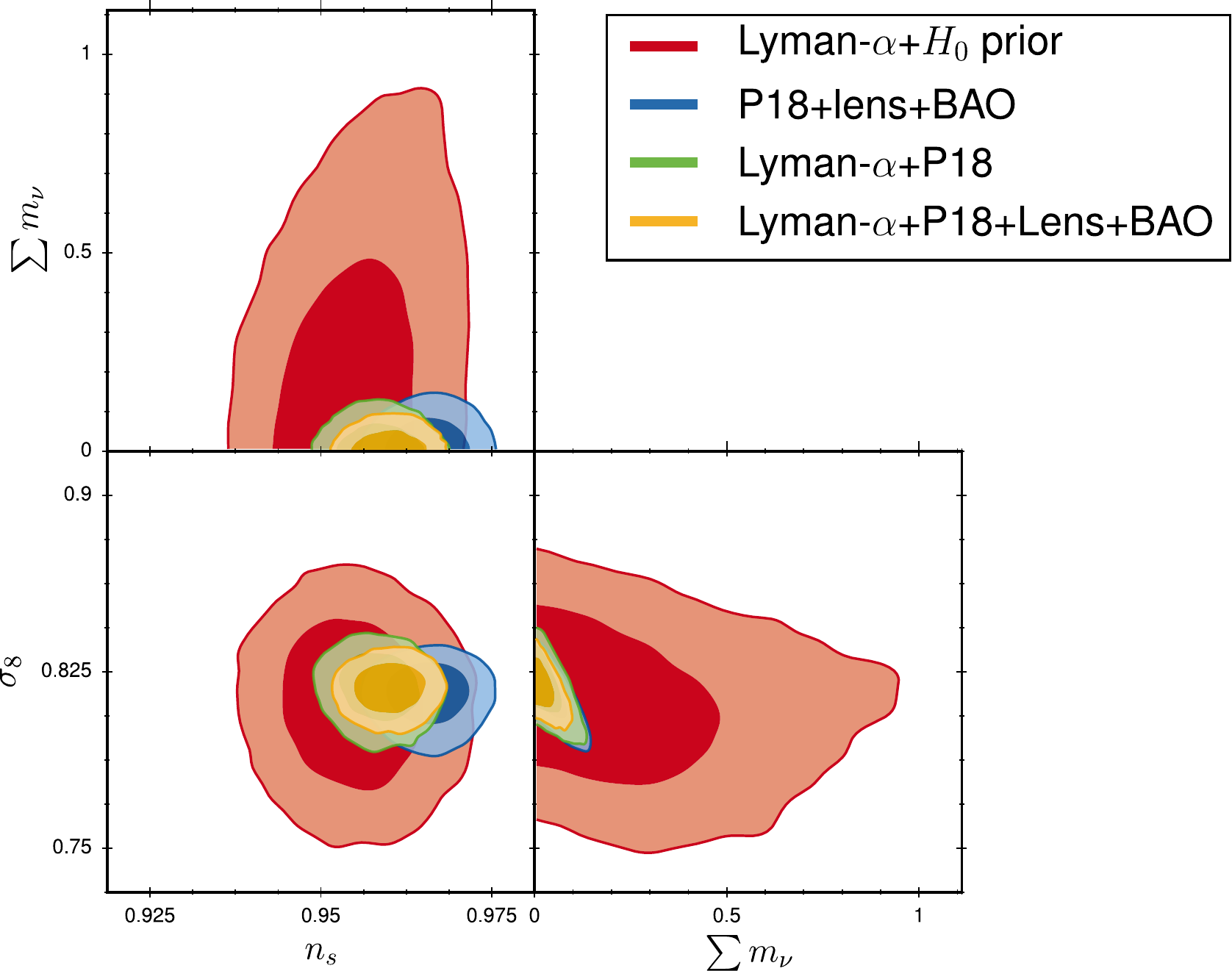,width = .9\textwidth}\\
\caption{\label{fig:Lya_vs_base_vs_full_vs_fullBAOlens} Bayesian marginalized 2D posteriors in the sub-space \{$\sum m_\nu$, $\sigma_8$, $n_s$\} for the \lcdmnu model and various combinations of CMB, BAO and \lya data. We show the 68.3\% $(1\sigma)$ and 95.4\% $(2\sigma)$ limits.}

\end{center}
\end{figure}

Before combining the \lya data with CMB data, we  compare the results from the \lya 
DR14 flux power spectrum (see table~\ref{tab:results_lya_alone}) and from Planck 2018 (see table~\ref{tab:results_Planck}) with minimal assumptions on the cosmological model, i.e. in the framework of the \lcdmnu model with a free value of $\sum m_\nu$. 
The common free parameters in the \lya and Planck likelihoods are the primordial spectrum amplitude and spectral index $\{\sigma_8 \mathrm{~or~} A_s, n_s, \}$, the fractional density of matter $\Omega_m$, and possibly the neutrino mass $\sum m_\nu$. There are two more common parameters $\Omega_b$ and $H_0$, but the \lya data are so weakly sensitive to them that we fixed $\Omega_b$ and imposed an $H_0$ prior that guarantees agreement with Planck.

\begin{figure}[t]
\begin{flushleft}
\epsfig{figure=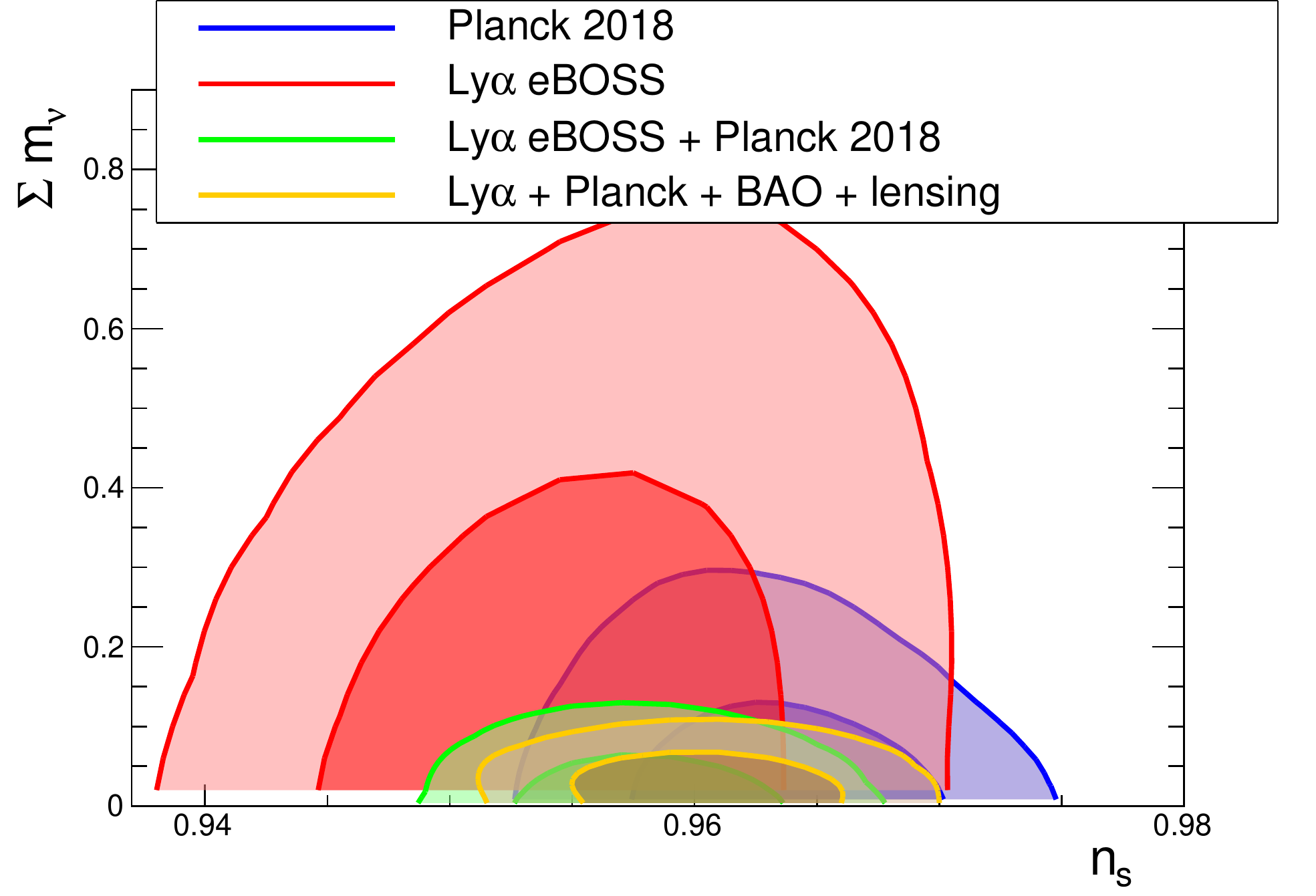,width = .48\textwidth}
\\
\epsfig{figure=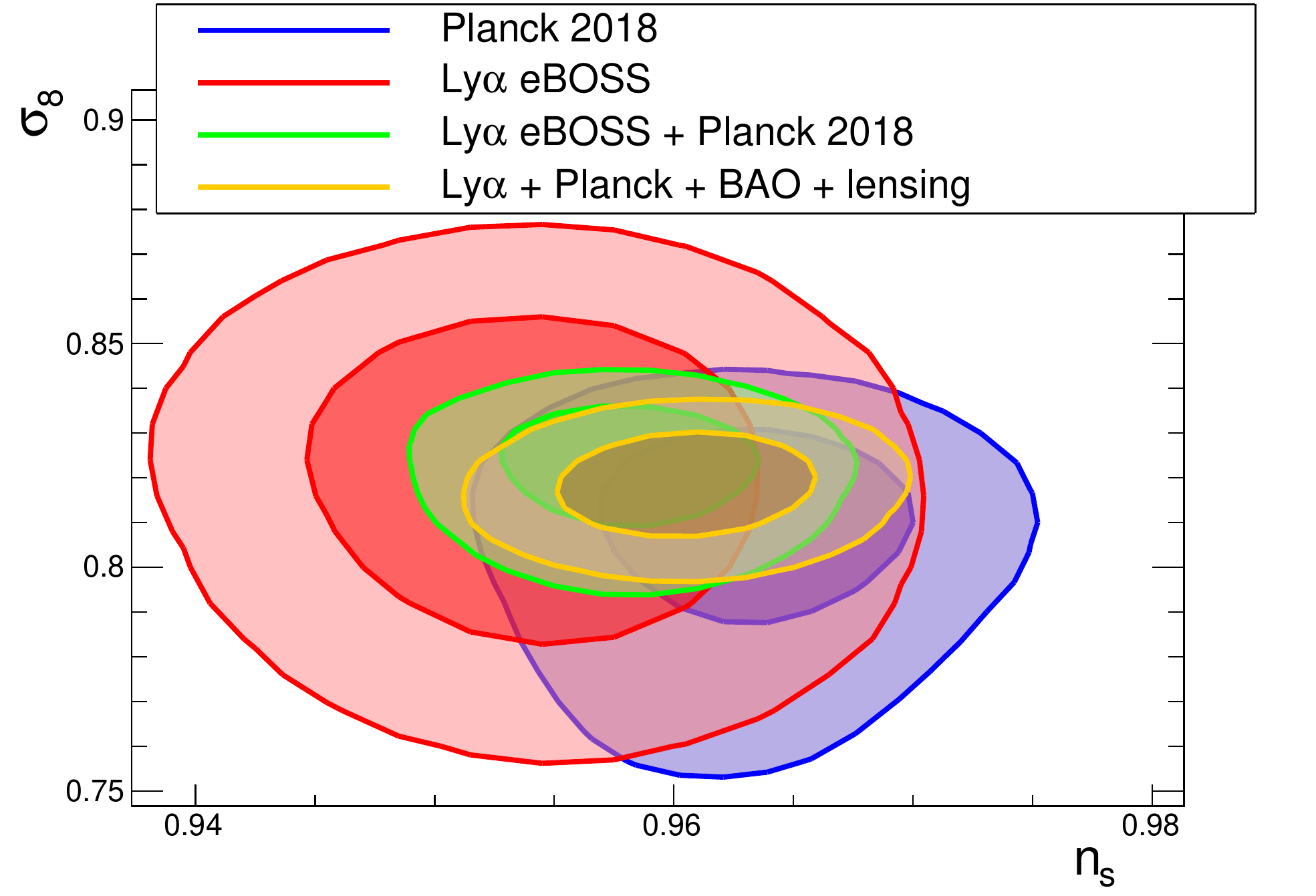,width = .48\textwidth}
\epsfig{figure=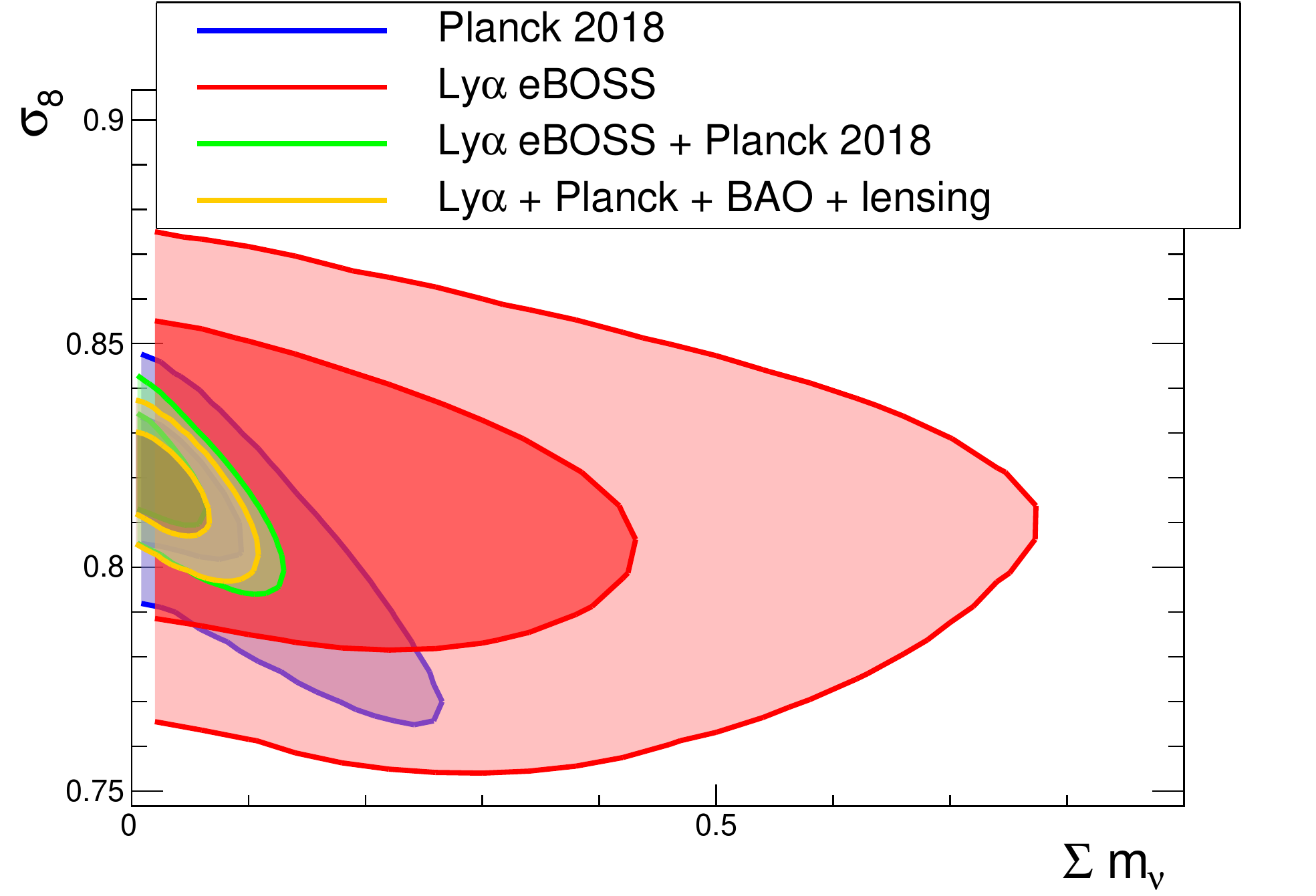,width = .48\textwidth}
\\
\end{flushleft}
\begin{center}
\caption{Frequentist 2D isocontours in the sub-space \{$m_\mathrm{tot} = \sum m_\nu$, $\sigma_8$, $n_s$\} for the \lcdmnu model and various combinations of CMB, BAO and \lya data. We show the $68\%$ and $95\%$ contours.}
\label{fig:Contour_2D_Freq}
\end{center}
\end{figure}

For all common parameters but one, we find  excellent agreement between the confidence bounds derived from \lya data and CMB data. This can be checked directly from  tables \ref{tab:results_lya_alone} and \ref{tab:results_Planck}, or visually by comparing two-dimensional contours in the \{$\sum m_\nu$, $\sigma_8$, $n_s$\} plane in figures \ref{fig:Lya_vs_base_vs_full_vs_fullBAOlens} and \ref{fig:Contour_2D_Freq}. In the Bayesian case, one should look at figure \ref{fig:Lya_vs_base_vs_full_vs_fullBAOlens} and compare red contours (\lya) with blue contours (P18+lens+BAO). In the frequentist case, figure~\ref{fig:Contour_2D_Freq} shows again the \lya contours in red and the P18 contours in blue. For instance, for $n_s$, the \lya and P18+lens+BAO bounds are compatible at the 1.4$\sigma$ level.

\begin{figure}[t]
\begin{center}
\epsfig{figure=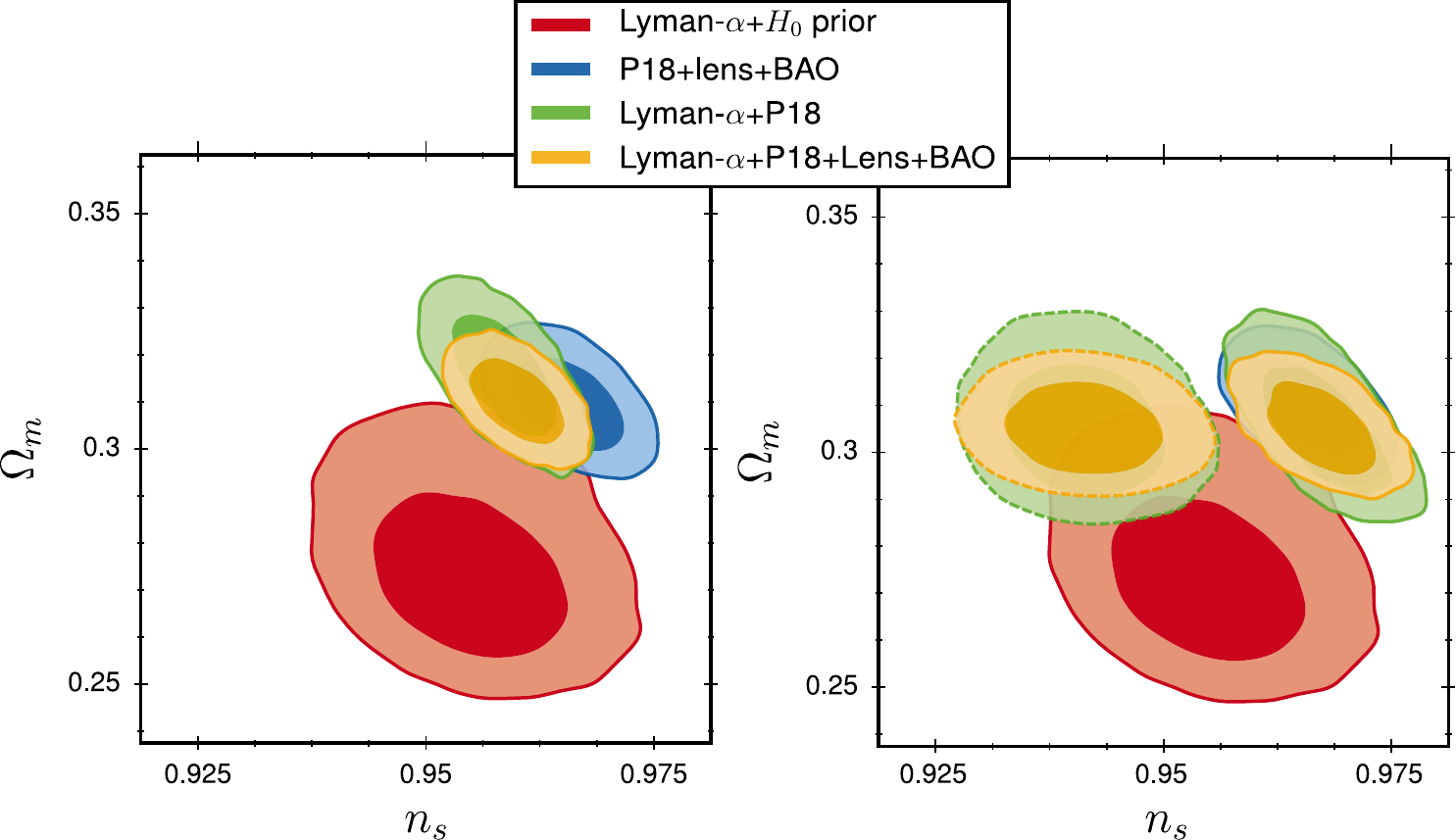,width = .9\textwidth}\\
\caption{\label{fig:Lya_vs_base_vs_full_vs_fullBAOlens_tension} 
Bayesian marginalized 2D posteriors in the sub-space \{$\Omega_m$, $n_s$\} assuming various cosmological models and combinations of CMB, BAO and \lya data. We show the $68.3\% (1\sigma)$ and $95.4\% (2\sigma)$ limits. {\bf Left:} \lcdmnu model, showing a mild tension. {\bf Right:} \lcdmnu model with two independent tilts for the Planck and \lya likelihoods.}
\end{center}
\end{figure}

We find a mild tension between the $\Omega_m$ values derived from CMB data \mbox{($\Omega_m \sim 0.31$)} and \lya data ($\Omega_m \sim 0.27$). The tension is present with both methodologies, with or without massive neutrinos, and with respect to both the P18 and P18+lens+BAO datasets. For instance, in the \lcdmnu model, the Bayesian (resp. frequentist) analysis returns a $2.5\sigma$ (resp. $3.6\sigma$) tension between the $\Omega_m$ values derived from the BOSS DR14 flux spectrum and the P18+Lens+BAO combination. This tension is displayed in the $\{\Omega_m, n_s\}$ plane in the left panel of figure~\ref{fig:Lya_vs_base_vs_full_vs_fullBAOlens_tension}.

Figure ~\ref{fig:Lya_vs_base_vs_full_vs_fullBAOlens_tension} (left panel) shows that for a fixed value of $\Omega_m$ around $0.31$, this mismatch could instead be interpreted as a mild tension on $n_s$. This is reminiscent of a similar tension on $n_s$ found with previous \lya data from BOSS DR9 discussed in \cite{Palanque-Delabrouille:2015pga}. When fitting the parameters of the \lcdm model or its extensions to \lya data, $n_s$ and $\Omega_m$ are always anti-correlated, because they both affect the overall slope of the flux power spectrum in the same direction. Therefore the tension on $\Omega_m$ in the present version of the data set and likelihood is likely to have the same origin as the tension on $n_s$ in the previous version: whether the tension is interpreted as one on $n_s$ or as one on $\Omega_m$ strongly depends on the modeling of the data and its systematics. In any case, since the analysis presented in this work is based on the most up-to date data set and on the most advanced systematic modeling of the BOSS and eBOSS flux power spectrum, we will concentrate on the $\Omega_m$ tension and its possible origins.

We stress that the \lya results are nicely consistent with those from weak lensing (WL) surveys. Over the past years, there has been a mild but persistent tension between likelihood contours in the $(\Omega_m, \sigma_8)$ plane inferred from Planck data and from WL surveys, when assuming either a \lcdm or \lcdmnu cosmology. This is commonly referred to as the \mbox{\tquote{$\sigma_8$ tension}}, although $\Omega_m$ is also involved. The tension is actually best seen when quoting results on the combination $S_8 \equiv \sigma_8 (\Omega_m/0.3)^{0.5}$ which is orthogonal to a direction of degeneracy in the WL posteriors.


\begin{figure}[t]
\begin{flushleft}
\epsfig{figure=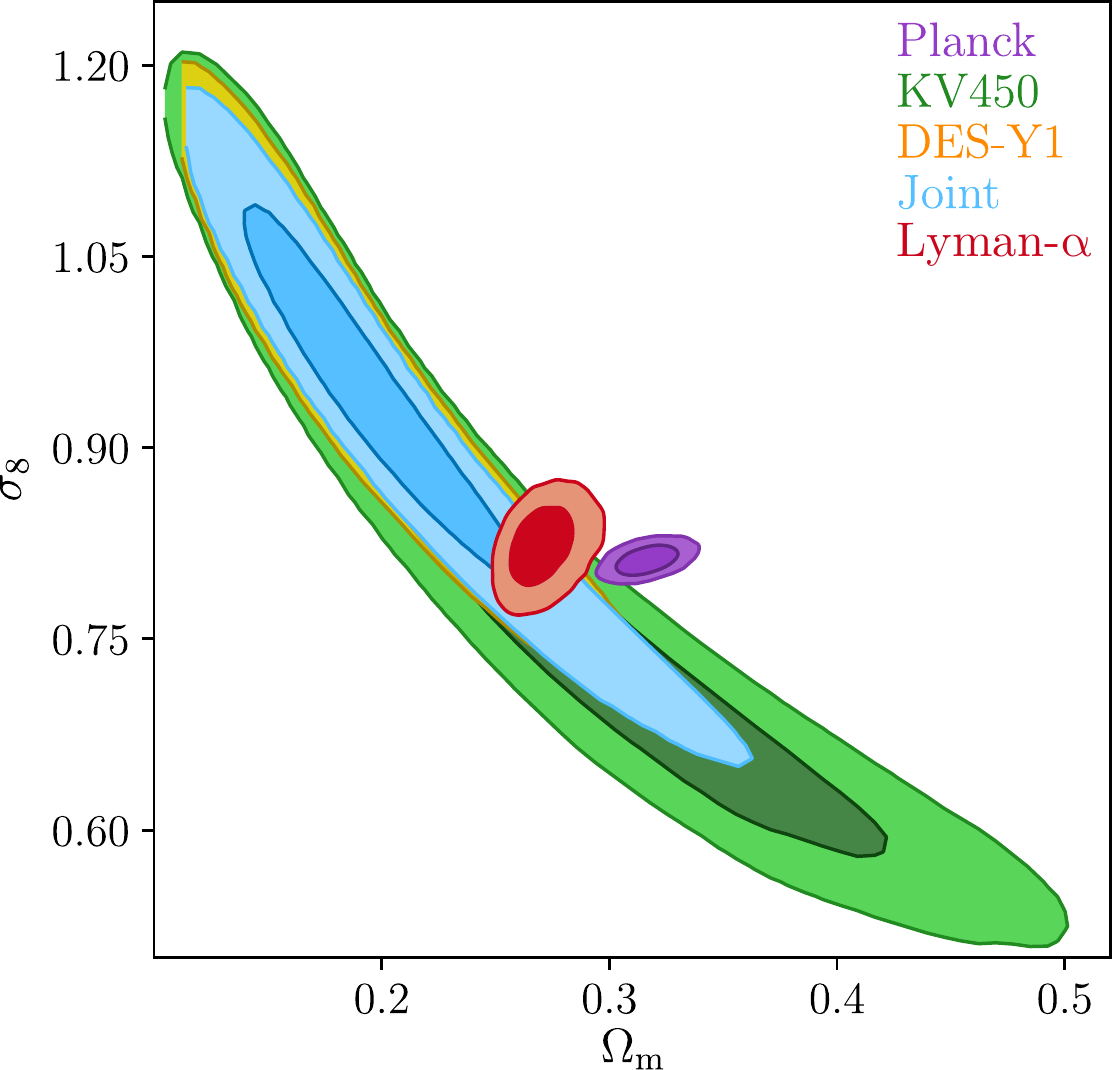,width = .475\textwidth}
\epsfig{figure=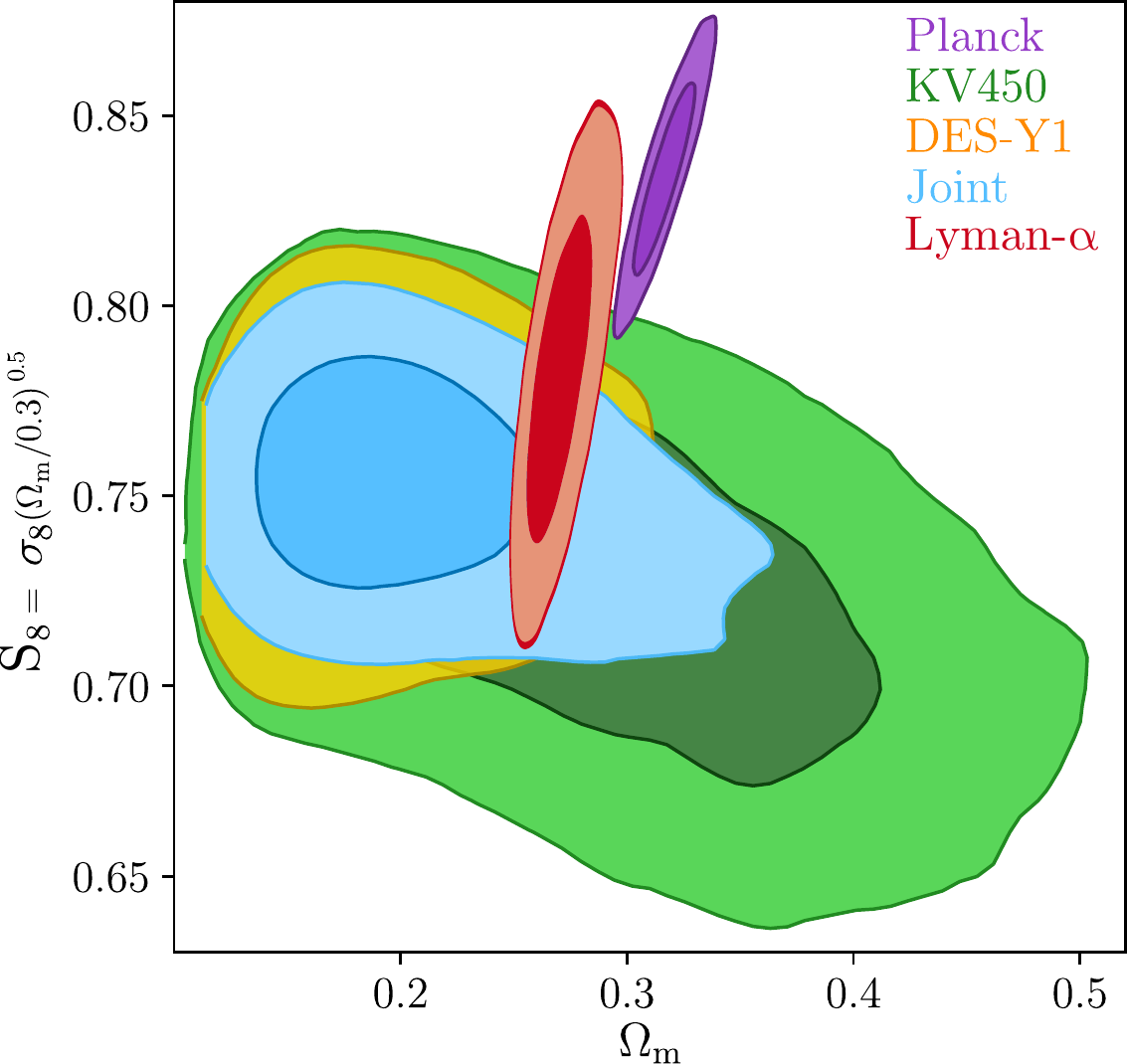,width = .485\textwidth}\\
\caption{\label{fig:Lya_vs_LSS_lens_tension} 
Comparison of the $\{\sigma_8,\Omega_m\}$ and $\{S_8,\Omega_m\}$ planes for Planck, Weak Lensing surveys (DES, KiDS+VIKING), and \lya data, where $S_8 \equiv \sigma_8 \left(\Omega_m/0.3\right)^{0.5}$. The tension between \lya and CMB data is here best described as a tension in $\Omega_m$. The COSEBI-based redshift-recalibrated analyses for DES-Y1\protect\footnotemark, KV450, and their joint constraint are taken from Asgari et al.~\cite{Asgari:2019fkq}.}
\end{flushleft}
\end{figure}
\footnotetext{Instead of using the DES-Y1 redshift distributions \cite{Troxel:2017xyo}, these new analyses use photometric redshifts from COSMOS-2015 \cite{Joudaki:2019pmv,Asgari:2019fkq,Laigle:2016jxn}. Note also that we are considering a flat \lcdm model in agreement with the cited analyses.}
In Figure~\ref{fig:Lya_vs_LSS_lens_tension}, we show the contours of the new BOSS + eBOSS DR14 \lya data (combined again with a $H_0$ prior) in the $(\Omega_m, \sigma_8)$ and $(\Omega_m, S_8)$ planes, compared with those from one of the most recent joint analyses \cite{Asgari:2019fkq} of several WL data sets (DES-Y1 \cite{Drlica-Wagner:2017tkk,Zuntz:2017pso}, KV450 \cite{Wright:2018nix}), and finally compared with Planck contours, assuming in each case a \lcdm cosmology. The left panel of Figure~\ref{fig:Lya_vs_LSS_lens_tension} shows that the $\sigma_8$ tension can be equally well interpreted as an $\Omega_m$ tension. While the CMB versus WL tension is strongest in the $S_8$ direction, the CMB versus \lya tension is strongest in the $\Omega_m$ direction. It is striking to see that WL and \lya data, which are two late time probes of a similar range of scales, agree with each other at the $1\sigma$ level, while they are both in tension with the Planck best-fit \lcdm model at the 2.5$\sigma$ to 3.6$\sigma$ level. 

We  performed a systematic search of the  origin of this tension, investigating  possible sources of systematics both at the level of the modeling of the \lya 1D flux power spectrum and at the level of  the \lya data analysis. In table~\ref{tab:test_Delta_chi2}, we summarize in each case the improvement in the consistency of the \lya and the CMB likelihoods, which we quantify by the change in the  $\chi^2$ difference between separate and combined data sets. A negative $\Delta \chi^2$ indicates an improved agreement, whereas a positive $\Delta \chi^2$ shows an enhanced tension. 
\begin{table}[t]
\begin{center}
\begin{tabular}{lc|lc}
\hline\hline
Configuration & $\Delta \chi^2$ &  Configuration & $\Delta \chi^2$ \\
 \hline \\[-10pt]
$ 0.0010 < k <0.017$ &  $ -0.2$  & DLA model (Rogers) & $0.0 $ \\
$ 0.0025 < k < 0.020 $ & $ -4.3 $  & Alternative splicing (1)   & $-2.1 $ \\
$ 2.1 <z < 4.5$ & $ +2.4 $  & Alternative splicing (2)   & $-0.6 $ \\
 $ 2.1 <z < 4.3$ & $ +8.4 $ & Top hat prior on $H_0$  & $+5.8 $ \\
 $ 2.3< z < 4.7 $ & $ +3.4 $  & Two independent $n_s$  & $ -13.8 $ \\
 $ 2.5 < z < 4.7$ &  $ +2.5 $  & Running of $n_s$  & $ - 8.2 $ \\
\hline
\end{tabular}
\caption{\label{tab:test_Delta_chi2} Tests performed to investigate possible sources of improvement of the consistency between \lya and Planck likelihoods. Improvement is quantified by $\Delta \chi^2$: the  $\chi^2$ difference between separate and combined data sets.  }
\end{center}
\end{table}
We first considered sub-samples of the \lya data to identify possible regions in $k$ or in $z$-space that would pull the fit away from the model preferred by CMB data. The results are reported in the left column of table~\ref{tab:test_Delta_chi2}. We  tested the impact of the smallest scales (cutting out $k>0.017\;{\rm s\;km^{-1}}$) as these are the most affected by our knowledge of the spectrograph resolution and our understanding of the noise power contribution (first row). We tested removing the large-scale modes (restricting to $k>0.0025\;{\rm s\;km^{-1}}$) instead, since these have the smallest statistical uncertainties and hence a large constraining power on the slope of the power spectrum (second row). We selected different redshift ranges, keeping only the lowest redshift bins (rows 3 and 4) that have better statistics, or instead the higher redshift ones (rows 5 and 6) that are less prone to systematics related to the measurement of the noise power. None of these tests yielded any significant change in the \lya best-fit cosmological values.  

In a second stage, we modified in several ways the model we use to fit the \lya data   (right column of table~\ref{tab:test_Delta_chi2}). The first test was using the correction suggested by~\cite{Rogers2018} to account for the incompleteness of the masking of the damped \lya systems (DLA) in the data, instead of the one from~\cite{McDonald2005} that was used in~\cite{Chabanier2019}. This had no impact, as shown in the first row. 
As detailed in~\citepy{}, the splicing technique produces a $k$-dependent bias caused by the change of splicing regime at a pre-determined pivot scale. The bias can be modeled by a broken line with a possible, albeit small, trend for some redshift evolution. We applied two new correction models to test whether the model we had opted for was responsible for the different slopes preferred by the \lya and the CMB data. 
The model  adopted in~\citepy{} includes a low-$k$ slope and a redshift evolution that are taken from the splicing study, and two free parameters: the high-$k$ slope and the offset at the pivot scale. 
Alternative splicing~1 is a model with no redshift dependence, a high-$k$ slope fixed to 0 and two free parameters:  the low-$k$ slope and the offset. Alternative splicing~2 has no redshift dependence, but allows for variation in both low-$k$ and high-$k$ slopes as well as a free offset. Neither splicing model yields a notable change on $\Delta\chi^2$. Finally, we also tested fitting the \lya data with a loose flat prior on $H_0$ between 55 and $80\;{\rm km\;s^{-1}\;Mpc^{-1}}$, instead of the usual Gaussian prior of $H_0=67.3\pm1.0\,{\rm km\,s^{-1}\,Mpc^{-1}}$. This had negligible impact on the result, as expected from the earlier studies of \cite{Palanque-Delabrouille2015}.  The last two lines of table~\ref{tab:test_Delta_chi2}, assuming two distinct values of $n_s$ on large and small scales or a running of $n_s$, refer to the study we describe in detail in the next two sections, and which leads to a notable improvement on the agreement between \lya and CMB data.

We cannot exclude the possibility that the tension between the value of $\Omega_m$ preferred by CMB and \lya data originates from yet another unidentified systematics (or imperfection in the computation of the \lya likelihood). Indeed, with statistical uncertainties on the data points down at the percent level, results on cosmological parameters are now hitting the systematics floor. Despite the great care that went into their modeling, instrumental features -- such as correction of spectrograph resolution and subtraction of noise power -- affect the lowest redshift bins and the smallest scales at a level comparable to the statistical uncertainties. Uncertainties also arise on the simulation side, mostly in relation to the use of the splicing technique mentioned above. Although the bias induced by this approach was measured to be small, and although we mitigate the risk of an imperfect modeling by marginalizing over the parameters that correct for the impact of splicing, a residual bias on the large-scale correction would affect the slope of the 1D flux power spectrum and could be responsible for the observed tension. 

\subsection{Combining  CMB and \lya data \label{sec:lyacmb}}

\begin{table}[tp]
\begin{center}
\begin{tabular}{lcccc}
\hline\hline
&\multicolumn{2}{c}{Frequentist} & \multicolumn{2}{c}{Bayesian} \\
& P18 + \lya& P18 + \lya& P18 + \lya & P18 + \lya \\
&  &+lens. +BAO & & +lens. +BAO \\
 \hline \\[-10pt]

$T_0$ (z=3) {\scriptsize($10^3$K)} & $9.7\pm1.7$  & $9.8\pm2.0$  & $9.5\pm1.8$  & $9.5\pm1.8$ \\[2pt]
$\gamma$   &  $0.69\pm0.10$ & $0.68\pm0.11$   & $0.71\pm0.10$ & $0.71\pm0.10$\\ [2pt]
$\sigma_8$ &  $0.825 \pm 0.006$     & $0.819\pm 0.008$  & $0.818 \pm 0.010$     & $0.818\pm 0.007$\\[2pt]
$n_s$ &  $0.958\pm0.003$    &  $0.961 \pm 0.003$  &  $0.959\pm0.004$    &  $0.960 \pm 0.003$ \\[2pt]
$\Omega_m$  &  $0.311\pm0.006 $     &  $0.308\pm0.006$ &  $0.316\pm0.009 $     &  $0.310\pm0.006$ \\[2pt]
$\sum \! m_\nu$~{\scriptsize(eV , 95\% CL)} & $<0.099$& $< 0.089$ & $<0.099$& $< 0.074$ \\[2pt]

\hline
\end{tabular}
\caption{\label{tab:results_lya+Planck} Preferred astrophysical and cosmological parameter values (68.3\% confidence level) for the \lcdm + $m_\nu$ model, for combined \lya, CMB and BAO data.}
\end{center}
\begin{center}
\begin{tabular}{lcccc}
\hline\hline
&\multicolumn{2}{c}{Frequentist} & \multicolumn{2}{c}{Bayesian} \\
& P18 + \lya& P18 + \lya& P18 + \lya & P18 + \lya \\
&  &+lens. +BAO & & +lens. +BAO \\
 \hline \\[-10pt]

$T_0$ (z=3) {\scriptsize($10^3$K)} & $7.6\pm1.9$& $7.6\pm 1.8 $ & $8.2\pm1.6$ & $8.2 \pm 1.6$\\[2pt]
$\gamma$   &  $0.88\pm0.13$ & $0.88 \pm 0.08 $ & $0.90\pm0.12$ & $0.89 \pm 0.12$\\ [2pt]
$\sigma_8$ & $0.824\pm 0.008$& $0.820\pm 0.008 $ & $0.814\pm 0.010$ & $0.818 \pm 0.008$\\[2pt]
$n_s$(Planck) & $0.965 \pm 0.004$& $ 0.968 \pm 0.004$ & $0.968 \pm 0.005$ &  $0.967 \pm 0.004$\\[2pt]
$n_s$(\lya) & $0.942 \pm 0.006$ & $ 0.942 \pm 0.005 $  & $0.941 \pm 0.006$ & $0.941 \pm 0.006$\\[2pt]
$\Omega_m$  & $0.304\pm0.010$ & $0.304\pm0.006$   & $0.305\pm0.009$ &  $0.305\pm0.006$ \\[2pt]
$\sum \! m_\nu$~{\scriptsize(eV , 95\% CL)} & $< 0.126$ & $<0.104$ & $< 0.109$ & $<0.087$\\[2pt]

\hline
\end{tabular}
\caption{\label{tab:results_lya+Planck_twons} Preferred astrophysical and cosmological parameter values (68.3\% confidence level) for the \lcdm + $m_\nu$ model, for combined \lya, CMB and BAO data, when introducing artificially two distinct $n_s$ value in the \lya and CMB likelihood.}
\end{center}

\begin{center}
\begin{tabular}{lcccc}
\hline\hline
&\multicolumn{2}{c}{Frequentist} & \multicolumn{2}{c}{Bayesian} \\
&\lya +P18& \lya +P18  & \lya +P18  & \lya +P18 \\
& & +lens. +BAO& &+lens. +BAO \\
 \hline \\[-10pt]

$T_0$ (z=3) {\scriptsize($10^3$K)} & $7.8 \pm 1.8$ & $8.0\pm1.8$ & $8.5 \pm 1.6$& $8.5 \pm 1.6$ \\[2pt]
$\gamma$   & $0.80 \pm 0.12$   & $0.80\pm0.12$ & $0.81 \pm 0.11$& $0.80 \pm 0.11$\\ [2pt]
$\sigma_8$  &   $0.825 \pm 0.007$   & $0.821\pm 0.007$ & $0.817 \pm 0.010$& $0.817 \pm 0.008$\\[2pt]
$n_s$  &  $0.962 \pm 0.003$   &  $0.962 \pm 0.003$ & $0.962 \pm 0.004$ & $0.962 \pm 0.004$\\[2pt]
$\Omega_m$  & $0.306 \pm 0.007$  &  $0.307\pm0.006$ & $0.310 \pm 0.009$ & $0.308 \pm 0.006$\\[2pt]
$\alpha_s$ & $-0.010 \pm 0.003$ & $ -0.010 \pm 0.003$ & $-0.010 \pm 0.004$& $-0.010 \pm 0.004$\\[2pt]
$\sum \! m_\nu$~{\scriptsize(eV , 95\% CL)} & $<0.105$ & $< 0.089$ & $<0.101$& $< 0.088$\\[2pt]

\hline
\end{tabular}
\caption{\label{tab:results_lya_nrun} Preferred astrophysical and cosmological parameter values (68.3\% confidence level) for the \lcdm + $m_\nu$ + $\alpha_s$ model, for combined \lya, CMB and BAO data.}
\end{center}
\end{table}

CMB and \lya data can be combined with different assumptions on the underlying cosmology.

First, one can adopt the point of view that the tension on $\Omega_m$ described in section \ref{sec:tension} is sufficiently small that it should not prevent us from combining the data sets while sticking to the  \lcdmnu model. In this case, the combined limits on cosmological parameters -- using either the frequentist or Bayesian methodology -- are presented in table~\ref{tab:results_lya+Planck}. We will come back to the discussion of the neutrino mass bounds found in this analysis in a dedicated section~\ref{sec:mnu}. Since the Planck data has more statistical weight, the $\Omega_m$ values in the combined fit are driven to Planck best-fit values, as shown in figure~\ref{fig:Lya_vs_base_vs_full_vs_fullBAOlens_tension}.  The other cosmological parameters, already in good agreement between their best-fit values for the two data sets taken individually, do not change significantly. While Planck data carry no information per se on the IGM thermal history, they can affect the best-fit values of the astrophysical parameters through their correlations with the cosmological terms. For instance, the anti-correlation between $\Omega_m$ and $\gamma$ -- and to a lesser extent between $n_s$ and $\gamma$ -- is causing  $\gamma$ to decrease compared to its value for the fit with \lya data alone, since both $\Omega_m$ and $n_s$ are increased. All correlation coefficients, and comments thereof, are provided in appendix~\ref{sec:appendixB}.

\FloatBarrier

Since for a fixed value of $\Omega_m$ around $0.31$, the tension could be interpreted as a tension in $n_s$, it is interesting to check whether the two data sets can be brought in better agreement by allowing the overall slope of the CMB and \lya flux power spectra to be independent of each other. To check this, we first perform an academic study in which we assume different values of $n_s$ in the CMB  and in the \lya likelihoods. Our results for that case are presented in table~\ref{tab:results_lya+Planck_twons}.

In this configuration, the global $\chi^2$ decreases by $\sim 13.8$ for P18+lens+BAO+\lya compared to the \lcdmnu fit, which represents an improved  compatibility between  \lya and P18+lens+BAO and data. The corresponding two-di\-men\-sion\-al $\{\Omega_m, n_s\}$ contours, displayed in the right panel of figure~\ref{fig:Lya_vs_base_vs_full_vs_fullBAOlens_tension}, now feature two distinct regions; one  (dashed lines) for $n_s$(\lya)  and another (solid lines) for $n_s$(Planck). The two values of $n_s$ are centered around $n_s=0.967$ for Planck and $n_s=0.941$ for \lya data. In the right panel of figure~\ref{fig:Lya_vs_base_vs_full_vs_fullBAOlens_tension}, the improved compatibility shows up in the fact that the red contour (\lya only) is now in better agreement with the green and yellow dashed contours (combined data set). 

Of course, the improved agreement occurs at the price of rather unphysical assumptions. This exercise is however not completely artificial, because the CMB and \lya data sets probe different fluctuations (photon perturbations or a tracer of baryons and CDM fluctuations) on different scales and at different times. Thus there could be many physical and sensible reasons for which the overall slope of the two observables are not correlated in the way predicted by the \lcdm or \lcdmnu models. 

First, the primordial power spectrum could have a different effective slope $\frac{d \ln {\cal P}_{\cal R}}{d \ln k}$ due to physical mechanisms taking place during inflation. For instance, a large curvature in the inflaton potential could produce a running of the spectral index, i.e. a continuous variation of $\frac{d \ln {\cal P}_{\cal R}}{d \ln k}$ with $k$ \cite{Kosowsky1995}, while a kink in the potential could lead to a feature in the primordial spectrum with different spectral indices on large and small scale \cite{Joy:2007na}. 

Second, the growth rate of dark matter could be reduced on small scales during radiation and/or matter domination, for instance due to small interactions between dark matter and other species or self-interactions in the dark matter sector, or by a small departure from Einstein gravity. Since baryons fall in the gravitational potential wells of dark matter, this reduction would propagate to the baryons and to the flux power spectrum. An example of a mechanism leading to a small reduction of the effective slope of the matter power spectrum $\frac{d \ln {\cal P}_\mathrm{m}}{d \ln k}$ is provided by the scattering dark matter model of \cite{Buen-Abad:2015ova}. Another case is that of $f(R)$ gravity, that leads to a scale-dependent linear growth factor with less growth on small scales~\cite{Motohashi:2010zz}.

In principle, a dedicated analysis would be needed in order to investigate up to which extent each of these models can reduce the mild tension between \lya and CMB data. In this work, we limit ourselves to the most studied among the previous models, featuring a running of the scalar index,  
$\alpha_s=\frac{d^2 \ln {\cal P}_{\cal R}}{d \ln k^2}=\frac{d n_s}{d \ln k}$, 
treated as constant over the range of scales probed by both CMB and \lya data. 

\subsection{\texorpdfstring{$\Lambda$CDM$\nu+\alpha_s$}{LCDMnu+alphas} cosmology \label{sec:running}}

On the theoretical side, the running of the spectral index is usually connected to the physics of inflation, but we should keep in mind that it could be seen as an effective parametrization of some of the other models described previously -- in particular, a negative $\alpha_s$ gives a reduction of the amplitude and of the effective slope of the small-scale matter power spectrum that could mimick a scale-dependent reduction of the linear growth factor. The simplest inflationary models predict that the running of the spectral index should be of second order in inflationary slow-roll parameters and therefore small, $|\alpha_s|\sim (n_s-1)^2\sim 10^{-3}$~\cite{Kosowsky1995}. Nevertheless, it is possible to accommodate a larger scale dependence of $n_s$ by adjusting the third derivative in the inflaton potential (see for instance \cite{Kobayashi:2010pz,McAllister:2008hb}). 

On the experimental side, recent CMB experiments have a mixed history of  null results and a-few-sigma detections of running of the scalar index. The final 9-year WMAP analysis found no evidence of running using WMAP alone, with $\alpha_s = -0.019\pm 0.025$ at 68\% CL, while the combination of WMAP data with the first data releases from ACT and SPT found a negative running at nearly the $2\sigma$ level with $\alpha_s = -0.022\pm 0.012$ \cite{Hinshaw2013}. The ACT 3-year release measured $\alpha_s = -0.003\pm 0.013$ when combining with WMAP-7~\cite{Sievers2013}. A negative running was detected at just over $2\sigma$ by SPT, $\alpha_s = -0.024\pm 0.011$~\cite{Hou2014}. The Planck 2018 results, 
 while roughly consistent with zero running of the scalar spectral index,  indicate a $\sim 1\sigma$ preference for negative running, $\alpha_s = -0.0041\pm 0.0067$. Finally, in the paper \citepy{}, we reported a $\sim 3\sigma$ tension on $n_s$ when we combined Planck 2015 and the DR9 BOSS \lya  measurement, which yielded $\alpha_s = -0.0117 \pm 0.0033$.

The analysis of \citepy{}, however, was  simplified:  the effect of running on the \lya likelihood was approximated as a change in the spectral index following the relation $n_s(k) = n_s(k_p) + \alpha_s \times \ln (k/k_p)$. In the present work, we performed dedicated simulations accounting for the full effect of running on the primordial spectrum, and added $\alpha_s$ to the list of parameters in the Taylor expansion of the flux power spectrum and in the \lya likelihood.
The result of our combined fit of CMB and \lya data for the $\Lambda$CDM$\nu+\alpha_s$ cosmology are presented in table~\ref{tab:results_lya_nrun}. The value of the spectral index reported here is defined at the pivot scale $k_*=0.05$~Mpc$^{-1}$. 

As in section~\ref{sec:LyaCMB}, the Bayesian and frequentist approaches provide consistent results. 
Allowing a running of $n_s$ improves the global fit to P18+lens+BAO+\lya by $\Delta \chi^2 \sim 8.2$ compared to the plain \lcdmnu model. This is a bit less than when floating two independent tilts, which yielded $\Delta \chi^2 \sim 13.8$. There are two reasons for this. First, the \lya data alone shows no preference for negative running, i.e. for a negative curvature in the shape of the flux power spectrum. Thus, although the  $\chi^2$ of the combined  fit decreases,  the \lya contribution to the total $\chi^2$ increases slightly. 
Second, in the model with running, negative values of $\alpha_s$ also suppress the fluctuation amplitude on the smallest and largest scales probed by Planck (i.e., in the regions of the Sachs-Wolfe plateau and Silk damping tail). This disfavors too low values of the running, as shown by the Planck-only bound $\alpha_s = -0.0041\pm 0.0067$. The fit with two independent $n_s$ values does not have this effect on CMB data and allows for a slightly stronger variation of the spectrum slope between CMB and  \lya scales, explaining the stronger decrease in the best-fit $\chi^2$. 

In a model with non-zero running, the \lya data are compatible with significantly larger values of $\Omega_m$ than in the \lcdmnu model. With a large negative running, the effective slope of the spectrum on \lya scales can be small (corresponding to a small effective $n_s$(\lya)), and thus $\Omega_m$ can be large. We obtain a detection of running at the $\sim 3\sigma$ level. With both Pl8+\lya and Pl8+\lya+lensing+BAO data sets, we find $\alpha_s = -0.010 \pm 0.004$, in agreement with the previous measurement of PY15. 

\subsection{Neutrino mass bounds \label{sec:mnu}}

\begin{figure}[t]
\begin{center}
\epsfig{figure=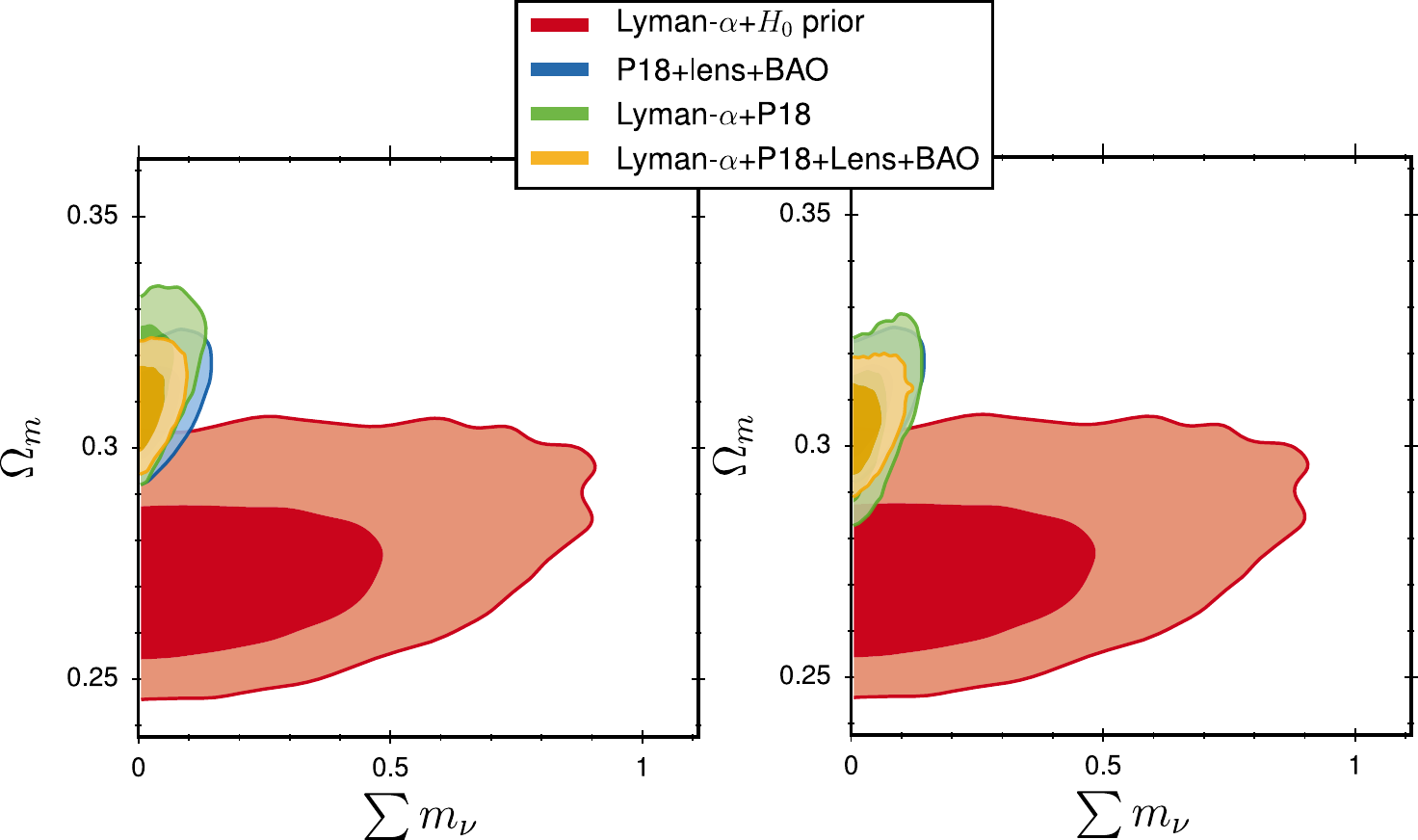,width = .9\textwidth}
\caption{\label{fig:omegam-mnu-bayesian} Bayesian marginalized 2D posteriors in the sub-space \{$\sum m_\nu$, $\Omega_m$\} assuming various cosmological models and combinations of CMB, BAO and \lya data. We show the $68.3\% (1\sigma)$ and $95.4\% (2\sigma)$ limits. {\bf Left:} \lcdmnu model, showing a mild tension. {\bf Right:} \lcdmnu model with two independent tilts in the Planck and \lya likelihoods.} 
\end{center}
\end{figure}

The combination of \lya and CMB data presents several advantages. CMB data alone is more sensitive to $\sum m_\nu$ than \lya data alone, through CMB lensing, the integrated Sachs Wolfe effect, and the measurement of the angular diameter distance to recombination \cite{Lesgourgues:2006nd,lesgourgues2013neutrino,PhysRevD.98.030001}. Combining the data sets helps in breaking degeneracies between cosmological parameters, such as between $\sum m_\nu$ and $\sigma_8$ (or $A_s$), $n_s$ and $\Omega_m$. This contributes to  further tightening  the constraint on $\sum m_\nu$. 

The constraint on $\sum m_\nu$ coming from \lya alone (with the $H_0$ prior) are included in table~\ref{tab:results_lya_alone}, and those from P18 or P18+lens+BAO in table~\ref{tab:results_Planck}. The joint bounds are presented in table~\ref{tab:results_lya+Planck} for the \lcdmnu model, table~\ref{tab:results_lya+Planck_twons} for the case with two independent spectral indices, and table~\ref{tab:results_lya_nrun} for the  $\Lambda$CDM$\nu+\alpha_s$ case. 

The joint \lya+P18+lens+BAO on $\sum m_\nu$ are loosened by a moderate amount (16\%) when switching from the \lcdmnu fit (frequentist bound of 0.89~meV at 95\%CL) to the fit with two free spectral indices (frequentist bound of 0.104~meV).
To understand this, we can look at 2D contours in the $(\Omega_m, \sum m_\nu)$ space for these two models, shown in figure~\ref{fig:omegam-mnu-bayesian}. The combined bounds are strongly influenced by the fact that \lya data remove a degeneracy between $\Omega_m$ and $\sum m_\nu$ in the CMB+BAO data. Thus the neutrino mass bounds must be discussed together with the mild $\sim 2\sigma$ tension between the values of $\Omega_m$ preferred by the two data sets. Since the tension is the strongest in the \lcdmnu model, the neutrino mass is the most constrained in this case. In the model with two spectral indices, the \lya data accommodate slightly larger values of $\Omega_m$, because they are partly compensated by lower values of $n_s$ at the level of the flux power spectrum. Thus the tension is relaxed and the neutrino mass bound gets a bit looser. The same trend is true in the $\Lambda$CDM$\nu+\alpha_s$ case, but the neutrino mass bounds  remain slightly tighter than in the case with the two $n_s$ values.

It is remarkable that all the neutrino mass bounds presented here fall within 20\% of each other, despite the very different assumptions on the scale dependence of the spectral index. Our results for $\sum m_\nu$ are thus found to be robust against various assumptions on the possible origin of the mild tension between \lya and Planck data when assuming a \lcdmnu cosmology. As a final result for neutrino masses, we choose to highlight the bounds coming from the most conservative of our two analyses based on a physical model, namely, the $\Lambda$CDM$\nu+\alpha_s$ analysis. We obtain $\sum m_\nu <0.11$eV  (resp. $< 0.09$eV) at the 95\% confidence level for \lya+P18 (resp. \lya+P18+lens. +BAO) data, both from the frequentist or the Bayesian approach. 

Given the results on neutrino oscillations (e.g.~\cite{deSalas2018} for a review), these results put  marginal tension on the inverted neutrino mass hierarchy scenario, which predicts a lower bound of 99.5~meV for~$\sum m_\nu$. Note however that an analysis adopting the \tquote{oscillation prior} $\sum m_\nu > 0.05$~meV would return a looser bound than our analysis assuming $\sum m_\nu > 0$. Thus, inverted hierarchy cannot be considered as disfavored at the 2$\sigma$ level by our data set.

We  also explored the impact of higher-resolution \lya data by including the XQ-100 flux power spectrum measured by~\cite{Yeche2017}. The fit to the extended \lya data set improves our sensitivity to neutrino masses, with tighter $3\sigma$ bounds and steeper likelihood profiles than in the case of DR14 \lya data alone. However,  XQ-100 has a very marginal preference for a non-zero neutrino mass, $\sum m_\nu=0.13\pm 0.18$~eV. Thus the combined bounds from \lya XQ-100 + DR14 data (including the $H_0$ prior) are not stronger than from DR14 data alone as one may have expected: we find  $\sum m_\nu<0.53$~eV for the joint bound. This trend was already noticed in~\cite{Yeche2017}, and shown for instance in their figure~10. The same is true when CMB data is added on top. The limit obtained from DR14 + XQ-100 + P18 data assuming a \lcdmnu cosmology, $\sum m_\nu<0.122$~eV, is slightly degraded compared to the one shown in table~\ref{tab:results_lya+Planck}, again in agreement with the study of~\cite{Yeche2017}.  

\subsection{Constraints on  \texorpdfstring{$\Lambda$WDM}{LWDM} model \label{sec:wdm}}

As explained in section~\ref{sec:lyasims}, the simulation grid can also be used to derive constraints on warm dark matter (WDM) in the form of thermal relics, since we  performed simulations with non zero values of $1\,{\rm keV}/m_X$. Here, we use the set of WDM simulations presented in~\cite{Baur2016} to provide results on  thermal relics and non-resonantly-produced (hereafter \tquote{NRP}) sterile neutrinos. We refer to~\cite{Baur2016} for extensive justification of this approach and we only summarize the main result here. 

As described by Dodelson and Widrow~\cite{DodelsonWidrow94}, a sterile neutrino population can be produced by non-resonant oscillations with the active sector in the early Universe (around $T \sim 100 \: \rm{MeV}$ for keV masses). This leads to a quasi-thermal distribution function for the sterile neutrinos, even if they never reached thermal equilibrium. Thanks to a symmetry in the evolution equations of linear perturbations, thermal WDM relics with a mass $m_X$ produce the same cut-off in the linear power spectrum as NRP sterile neutrinos with a mass $m_s$ inferred from the rescaling relation
\begin{equation}
\label{eq:MsMxrelation}
\frac{m_s}{3.90 \,\rm{keV}} = \left( \frac{m_X}{\rm{keV}} \right)^{1.294} \left( \frac{0.25 \times 0.7^{2}}{\Omega_{WDM}h^2} \right)^{1/3}\;.
\end{equation}
Since for this category of models, WDM simulations depend on the nature of WDM only through the input linear power spectrum~\cite{BLR09}, simulations with thermal relics can also be used to derive bounds on NRP sterile neutrinos.  Other classes, such as resonantly-produced sterile neutrinos, are of particular relevance. One cannot, however, constrain their mass through a simple rescaling of $m_X$. It requires a dedicated study that  is beyond the scope of this work.

\begin{figure}[t]
\begin{center}
\epsfig{figure=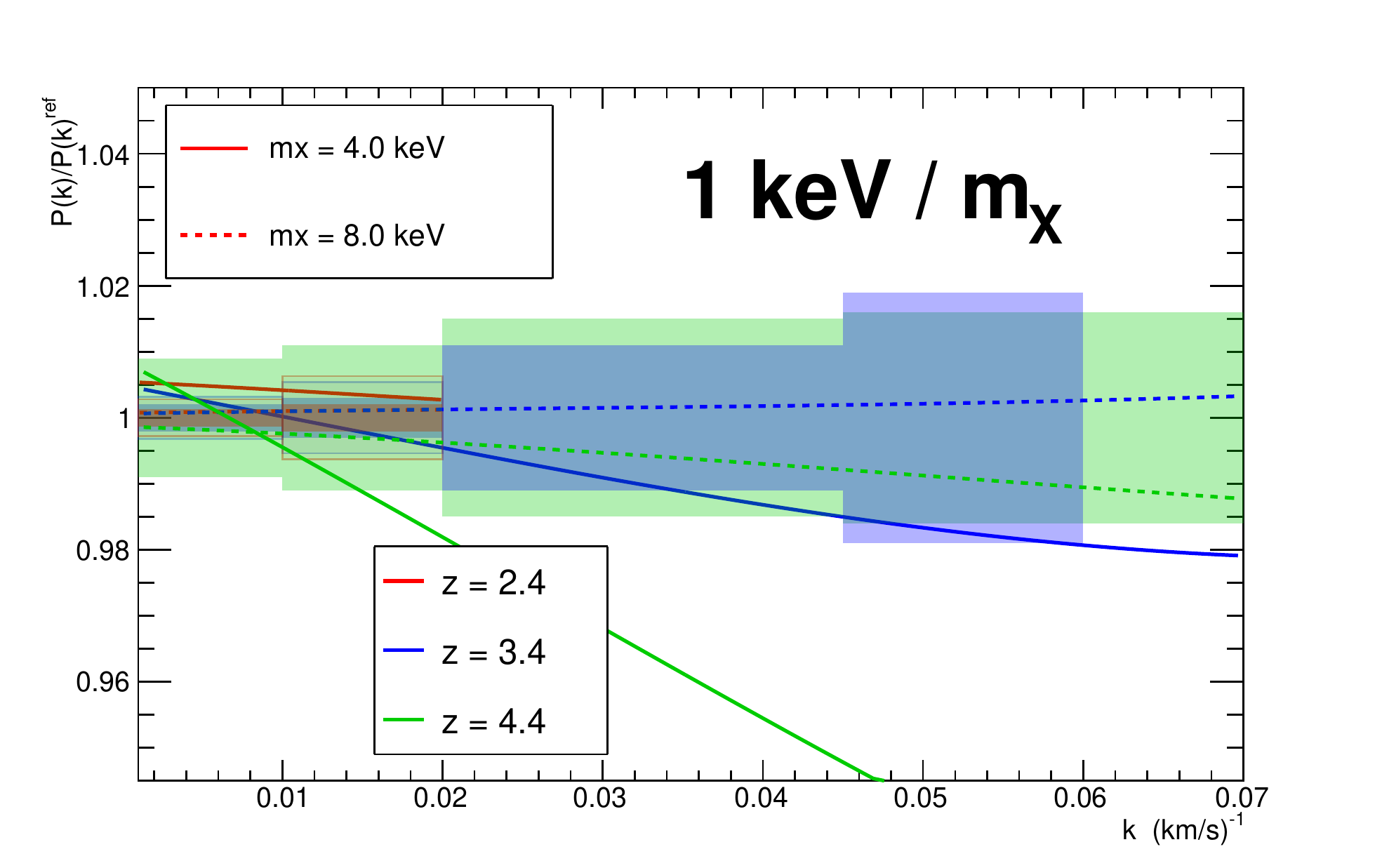,width = .7\textwidth}
\caption{Sensitivity of the eBOSS and XQ-100 data to WDM. The curves show the predicted deviation of the \lya flux power spectrum for a WDM  particle mass of 4.0~keV (solid) or 8.0~keV (dashed) relative to $\Lambda$CDM, in three redshift bins, when all other cosmological, astrophysical and nuisance parameters are kept fixed. The  boxes illustrate the data uncertainty  using the same code as for the curves. The z=2.4 bin only refers to eBOSS data and is bounded to $k<0.02\,\rm s\,km^{-1}$. Shaded boxes refer to statistical errors only, while larger clear boxes bounded by colored lines refer to statistical plus systematic errors. Systematics only significantly impact the total uncertainty for $k<0.02\,\rm s\,km^{-1}$ (eBOSS data) and  at small  redshift ($z<3.4$ typically). 
\label{fig:sensitivity}  }
\end{center}
\end{figure}

To constrain the mass of a WDM thermal relic, we use a \lya data set consisting of the eBOSS DR14 data combined with  the XQ-100 data from~\cite{Yeche2017}. The use of high-resolution data to constrain WDM has recently been subject to debate~\cite{Garzilli2015, Garzilli2019}. On the one hand, they are expected to be more sensitive to WDM as they better probe the scale range impacted by the WDM power suppression,  as illustrated in Figure~\ref{fig:sensitivity}. On the other hand, this  cut-off is partially degenerate with a similar effect  caused instead by a warm IGM. We here use the same model of the IGM thermal history as in the previous sections,  allowing, in addition, variations of $z_{\rm reio}$ as explained in section~\ref{sec:lyasims}. 

\begin{figure}[t]
\begin{center}
\epsfig{figure=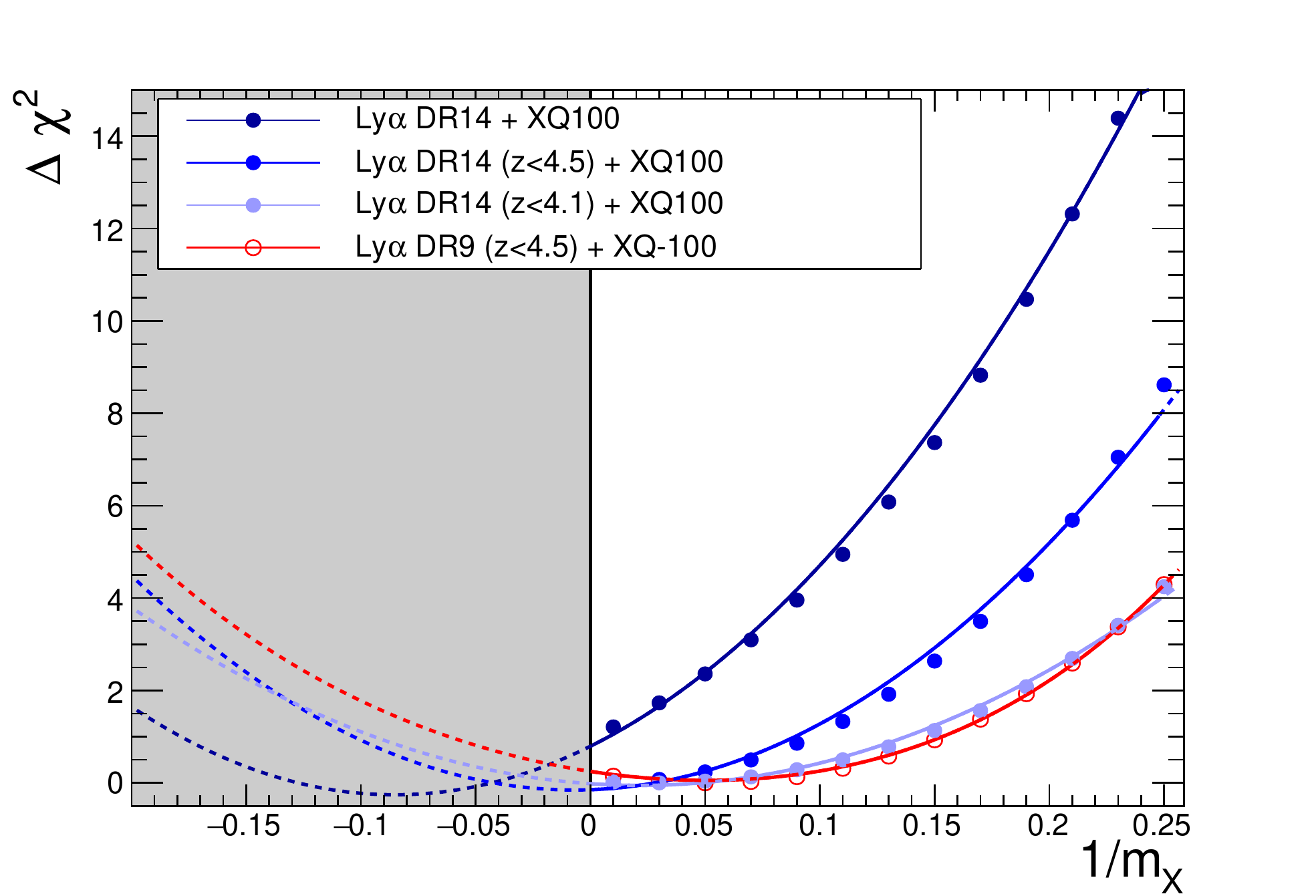,width = .7\textwidth}
\caption{\label{fig:plotchi2} $\Delta\chi^2$ profile as a function of $1\,{\rm keV}/m_X$ for the four configurations: \lya + XQ-100 (dark blue), \lya ($z\!<\!4.5$) + XQ-100 (blue),  \lya ($z\!<\!4.1$)+ XQ-100 (light blue), DR9 \lya ($z\!<\!4.5$) + XQ-100 (red with open circles). Each point shows the $\Delta\chi^2$ obtained from a profiling method, i.e. after minimization over all other free parameters. The curves are the result of a parabolic fit to the points, extrapolated into the negative region. } 
\end{center}
\end{figure}

The eBOSS DR14 + XQ-100 \lya data constrain the WDM mass to $m_X > 8.6$~keV (95\% CL).  Because non-linear structure growth tends to erase the power suppression caused by the free-streaming of relativistic particles, the sensitivity to $m_X$ is more prominent at high redshift. Thus, the high-redshift data are the ones with the largest constraining power, despite having the largest statistical uncertainties.  This feature is confirmed by the loosening of the bound as we restrict the data to lower redshifts. We obtain $m_X>5.3 \,{\rm keV}$ for $z<4.5$, and $m_X>3.9 \,{\rm keV}$ for $z<4.1$, both at 95\% CL. This trend is also clearly visible in figure~\ref{fig:plotchi2}, where  we show the $\chi^2$ profiles for the three configurations. The profile widens (hence the constraint loosens) as we go from all redshifts (dark blue) to $z<4.5$ (blue) then to $z<4.1$ (light-blue). 

The gain provided by the highest-redshift $z=4.6$ bin, however, is two-fold. It is due in part because the $z=4.6$ power spectrum probes structure growth in a more linear regime, but also because the minimum of the fit then occurs for $1\,{\rm keV}/m_X<0$. The Feldman-Cousins prescription~\cite{FC98} allows us to derive the limit in such a case, by computing the $\Delta\chi^2$ with respect to the $\chi^2$ at the limit of the physical domain, i.e. where $1\,{\rm keV}/m_X=0$. This method, however, leads to an artificially strong limit, since the $\chi^2$ profile in the physical ($m_X>0$) region is a very steep  function of $1\,{\rm keV}/m_X$. 
To be conservative, we hence consider as our main result the bound obtained in the eBOSS DR14 ($z\!<\!4.5$) + XQ-100 configuration: $m_X>5.3 \,{\rm keV}$ (95\% CL). We highlight the fact that removing the highest $z=4.6$ redshift bin is sufficient to ensure that the best-fit minimum be in the physical region. Further data restriction does not alter its location.  In table~\ref{tab:results_wdm}, we give the best-fit values of some of the main parameters with the two methodologies. We note an excellent agreement on all parameters.

\begin{table}[tp]
\begin{center}
\begin{tabular}{lcc}
\hline\hline
&{Frequentist} & {Bayesian} \\
&\lya (z<4.5)+XQ-100& \lya (z<4.5)+XQ-100 \\
 \hline \\[-10pt]

$T_0$ (z=3) {\scriptsize($10^3$K)} & $13.9\pm1.7$   & $12.2\pm2.0$ \\[2pt]
$\gamma$    & $1.09\pm0.13$ & $1.01\pm0.15$\\ [2pt]
$\sigma_8$  & $0.796\pm 0.020$  & $0.806 \pm 0.021$    \\[2pt]
$n_s$ &  $0.954\pm0.006$    &  $0.954 \pm 0.006$ \\[2pt]
$\Omega_m$  &  $0.265\pm0.008 $     &  $0.268\pm0.009$ \\[2pt]
$m_X$~{\scriptsize(keV , 95\% CL)} & $>5.3$& $>5.9$ \\[2pt]

\hline
\end{tabular}
\caption{\label{tab:results_wdm}  Preferred astrophysical and cosmological parameter values (68.3\% confidence level) for the WDM model, for combined \lya ($z<4.5$) and XQ-100 data.}
\end{center}
\end{table}

These new bounds are significantly tighter than the $4.2$~keV lower limit derived with BOSS DR9 ($z\!<\!4.5$) + XQ-100 by~\cite{Yeche2017} (red curve in figure~\ref{fig:plotchi2}). The improvement mostly comes from the improved statistical power of the eBOSS DR14 data compared to BOSS DR9, as is made clear by the comparison of the the limits obtained with the same  $z<4.5$ redshift range.

Because the IGM thermal history produces a cut-off in the flux power spectrum that is partially degenerate with the impact of a relativistic particle, we paid particular attention to the thermal modeling. The recovered $T_0(z)$, $\gamma(z)$ and mean flux are shown in appendix~\ref{sec:appendixA}. All three parameters are in perfect agreement with  typical observed ranges. We also checked the impact of a different modeling by imposing a central thermal history in agreement with the observations of~\cite{Becker2011} and allowing variations in the amplitude and density dependence of the UV heating rates (AMPL and GRAD parameters of the hydrodynamical code {\sc Gadget}). This only had a mild impact on the constraints, which  led to $m_X>4.7 \,{\rm keV}$ for eBOSS ($z\!<\!4.5$) + XQ-100. Finally, we give the correlation coefficients between the free parameters of the fit in appendix~\ref{sec:appendixB}. The $1/m_X$ parameter does not exhibit  significant correlation with any of the parameters describing the thermal history.

As our most robust bound on WDM, we therefore take the eBOSS ($z\!<\!4.5$) + XQ-100 configuration, marginalizing over the cosmological, astrophysical and nuisance parameters described in table~\ref{tab:astroparam}. This leads to $m_X>5.3 \,{\rm keV}$ (95\% CL), or equivalently to a constraint on the mass of a non-resonantly produced sterile neutrino $m_s>34\,{\rm keV}$ (95\% CL).

\section{Conclusions} \label{sec:conclusion}

In  this paper, we present an update of the constraints we derive on several cosmological parameters using \lya data, either  alone or in combination with  CMB and BAO data. Compared to the previous study of \citepy{}, we update both  large-scale and small-scale data sets: we use  the most recent 1D \lya flux power spectrum measured with the  DR14  BOSS and eBOSS data of the SDSS, as well as  the newest Planck 2018 data release. 

We perform  two statistical analyses in parallel: one based on a Bayesian and the other on a frequentist interpretation. The two approaches produce results that are in excellent agreement,  demonstrating the robustness of the study. In order to be conservative, we choose to always report as our final result the  largest (and hence weakest) bound, whether on the neutrino masses $\sum m_\nu$ or on the inverse of the mass of a thermal relic $1/m_X$.  

We find a mild tension between the values of $\Omega_m$ preferred by the \lya and the CMB data.  Interestingly, the \lya best-fit cosmological parameters are in very good agreement with current weak lensing constraints on ($\Omega_m,\sigma_8$).  \lya and weak lensing are two late-time probes of a similar range of scales, and they show a comparable level of tension with Planck $\Lambda$CDM model at the $2-3$~$\sigma$ level. 
Because $\Omega_m$ and $n_s$ have a similar impact on the \lya flux power spectrum, the small tension on $\Omega_m$ is likely to have the same origin as the mild tension on $n_s$  previously observed by \citepy{}. 

We performed comprehensive tests and did not identify a systematic effect in the data or in the analysis that would be the origin of this tension. However, we showed that it could be reduced by considering different scalar indices on CMB and \lya scales, such as produced by a running of the scalar index. 
We find a mild preference for a non-zero running of $n_s$ at the level of about $3\sigma$: ${\mathrm d}n_s/{\mathrm d}\ln k \sim - 0.010 \pm 0.004$. This detection is consistent with the previous study of PY15. It illustrates the small disagreement in the slopes of the power spectrum measured independently by BOSS/eBOSS and Planck. 

The free-streaming of massive neutrinos causes a step-like suppression in the power spectrum that is ideally probed by comparing the large-scale CMB  to the small scale \lya power spectra. The constraint on $\sum m_\nu$ thus comes from the measurement of $\sigma_8$ in \lya data, and from the correlation between $\sigma_8$ and $\sum \! m_\nu$ provided by CMB. The value of $\sigma_8$ is derived from the normalization of the 1D flux power spectrum, which is essentially unaffected by the inclusion or not of a variation in the slope of the power spectrum, as in the case of non-zero running. 

Combining BOSS and eBOSS \lya with Planck CMB data, we find an upper bound on $\sum m_\nu$ of 0.10~eV (95\% CL) for a $\Lambda$CDM model, which only loosens to $0.11$~eV when allowing for running. When further including CMB lensing and BAO, the limit tightens to $\sum m_\nu < 0.09$~eV, whether or not running is allowed. These limits improve slightly over those of PY15 and tend to favor the normal hierarchy neutrino mass scenario. 

WDM affects clustering compared to the CDM scenario by suppressing all power below a scale determined by the particle mass. Thanks to the much improved statistics, the DR14 BOSS and eBOSS data allow us to   improve the limit on WDM compared to previous publications. Using a combination of eBOSS ($z\!<\!4.5$) + XQ-100 \lya data, we constrain the mass $m_X$ of a thermal relic to  $m_X>5.3$~keV (95\% CL). It translates to a constraint on the mass $m_s$ of a non-resonantly-produced sterile neutrino of $m_s>34$~keV (95\% CL). 

The study presented in this work  probably provides one of the most stringent bound on $\sum m_\nu$ that current high-statistics medium-resolution data can provide. WDM constraints, less sensitive to low-redshift data  more prone to systematics, could still be further improved, for instance with additional high redshift data. We see two main paths for improvement in the near future. First, new-generation spectroscopic data from DESI, WEAVE, or 4MOST will soon be available. With a three to four-fold increase in quasar number density, it will be possible to further tighten the selection of the quasar spectra and  reduce the contamination from systematic biases. With DESI, the factor of two gain in resolution and the reduced noise will improve the measurement on small scales relevant for $\sum m_\nu$. The  extension of the DESI quasar selection to higher redshift is highly relevant to WDM. Secondly, improvements are also expected on the simulations side. The use of faster and less memory-intensive codes  will allow one to run large-volume hydrodynamic cosmological resolution with the required resolution without having to resort to  splicing techniques. Emulator-based  techniques can reduce the  uncertainties in the model interpolation and bring them at the sub-percent level. 
Combined, these improvements should allow us to enter the era where $\sum m_\nu$ is no longer constrained but measured.

\acknowledgments

We would like to thank Marika Asgari, Hendrik Hildebrandt and Konrad Kuijken for sharing the results of their recent Weak Lensing data analysis with us.
NS acknowledges support from the DFG grant LE 3742/4-1. JL is supported by the DFG grant LE 3742/3-1. MW, EA, NPD and CY acknowledge support from grant ANR-17-CE31- 0024-01 for the NILAC project. The Bayesian analysis was performed with computing resources granted by RWTH Aachen University under project jara0184. The hydrodynamical simulations were performed under PRACE allocations 2010PA2777, 2014102371 and 2012071264 and GENCI allocations t2013047004, t2016047706, A0030410313, A0050410586 on Curie thin and fat nodes at TGCC. 

\newpage
\appendix
\section{Thermal history}
\label{sec:appendixA}

\begin{figure}[htbp]
\begin{center}
\epsfig{figure= 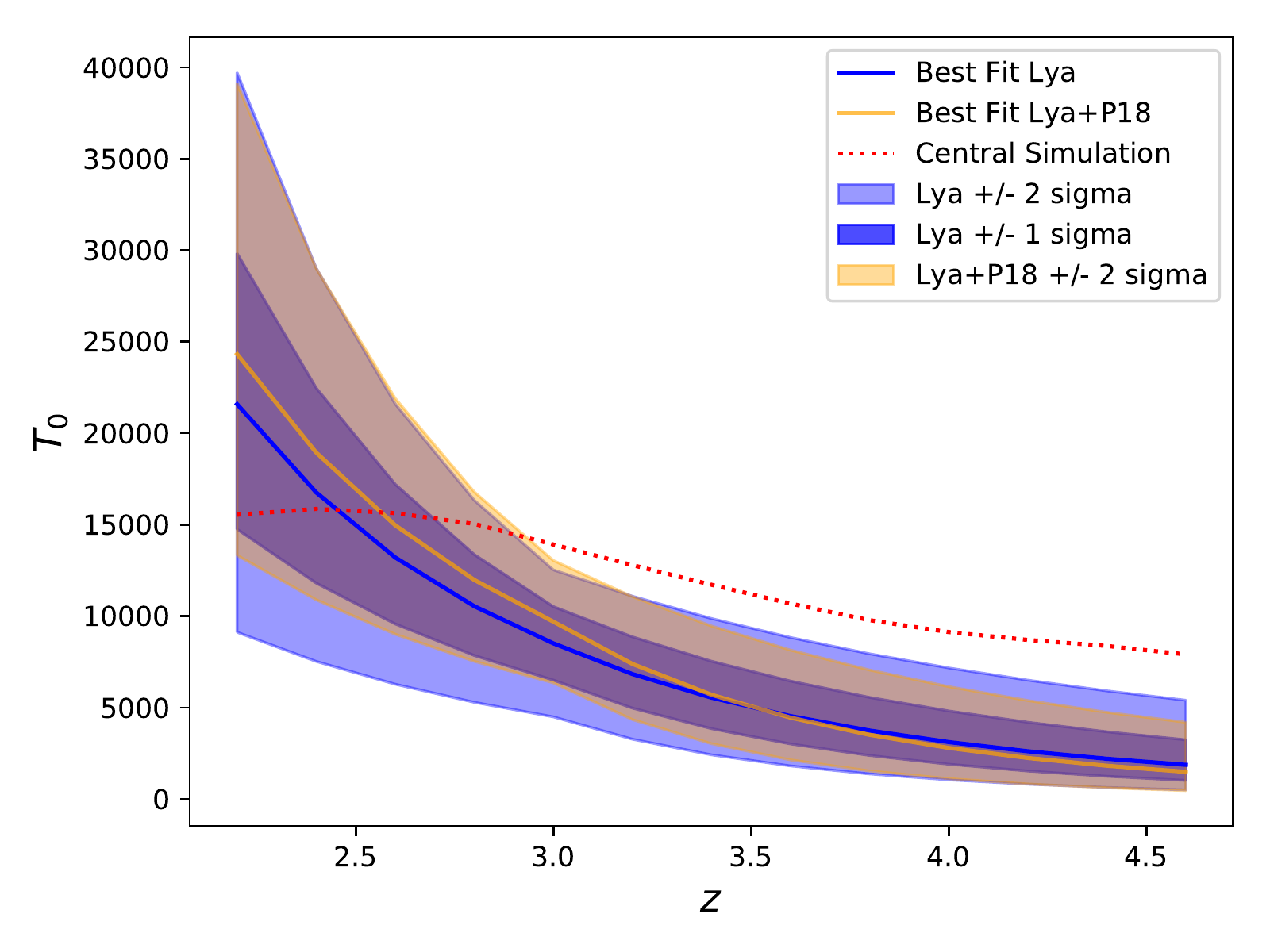,width = .49\textwidth}
\epsfig{figure= 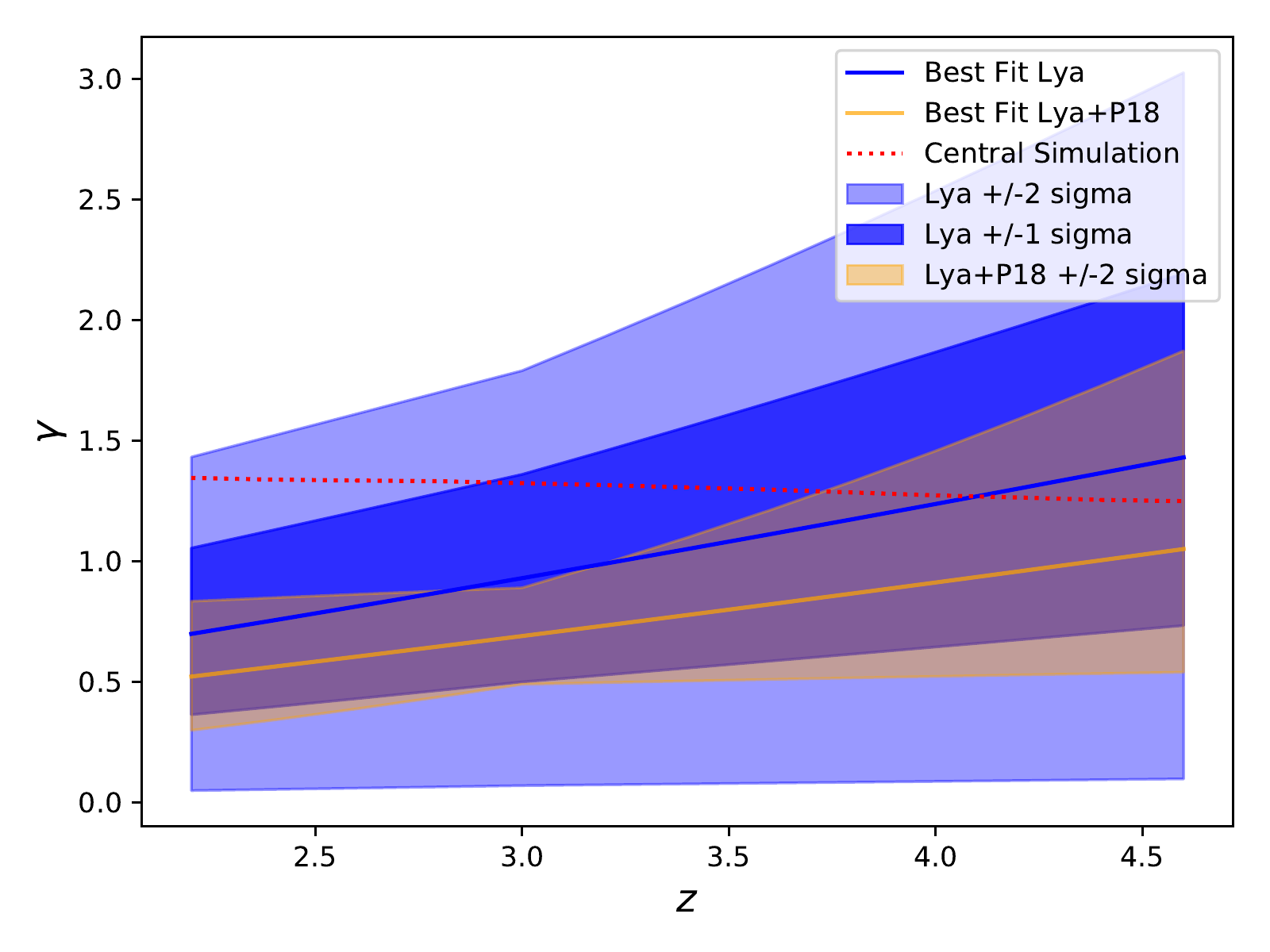,width = .49\textwidth}\\
\epsfig{figure= 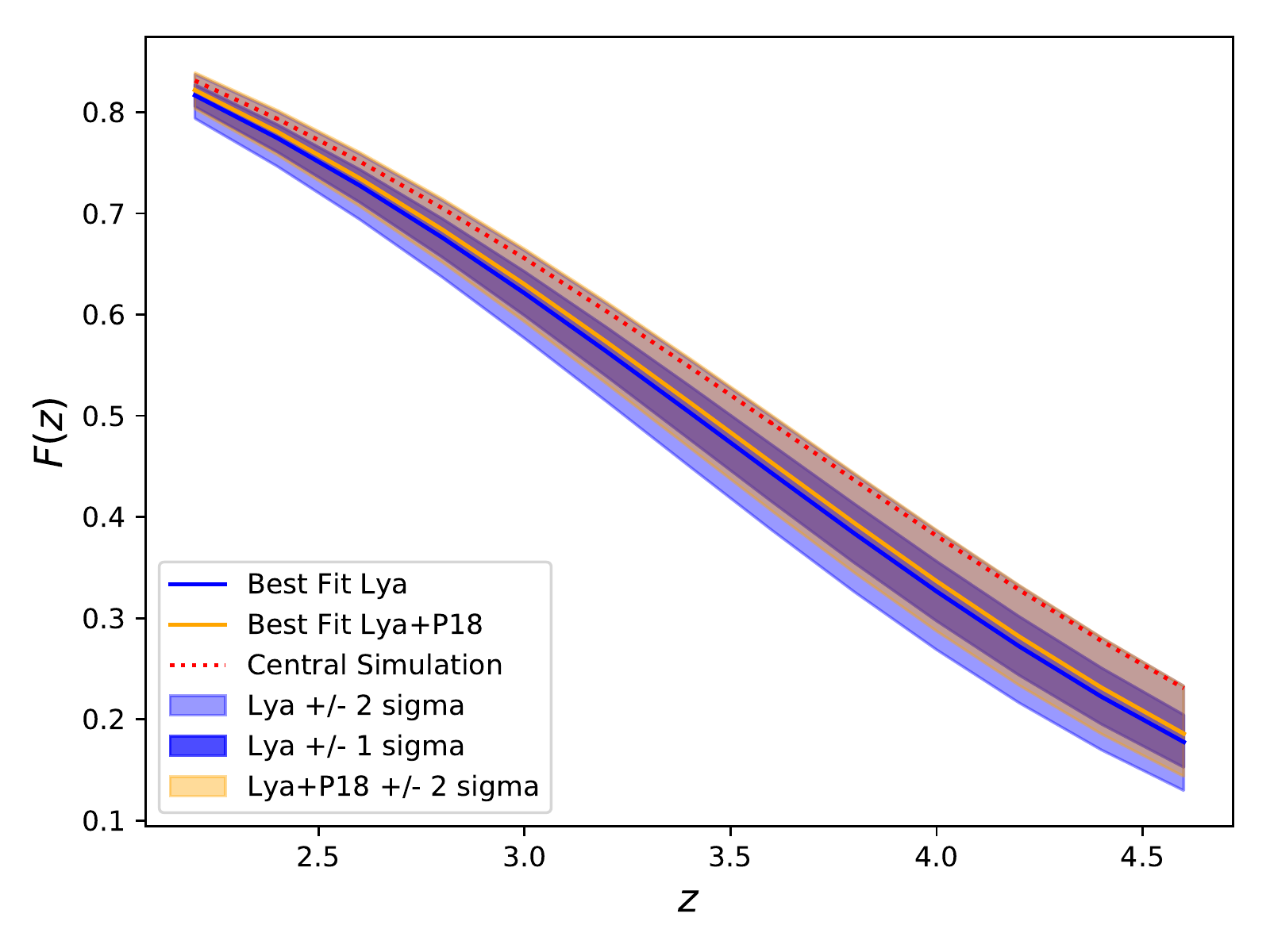,width = .49\textwidth}
\caption{Thermal history for the $\sum m_\nu$ fit. Blue contours for \lya data alone (1 and 2 $\sigma$ contours), yellow for \lya and Planck (2 $\sigma$ contour only). The dashed red curve illustrates the thermal history of the central simulation of the grid. \textbf{Top left:} temperature $T_0(z)$. \textbf{Top right:} Density dependence $\gamma(z)$ parameter. \textbf{Bottom:} mean flux $\overline{F}(z)$.  }
\label{fig:thermal_mnu}
\end{center}
\end{figure}

\begin{figure}[htbp]
\begin{center}
\epsfig{figure= 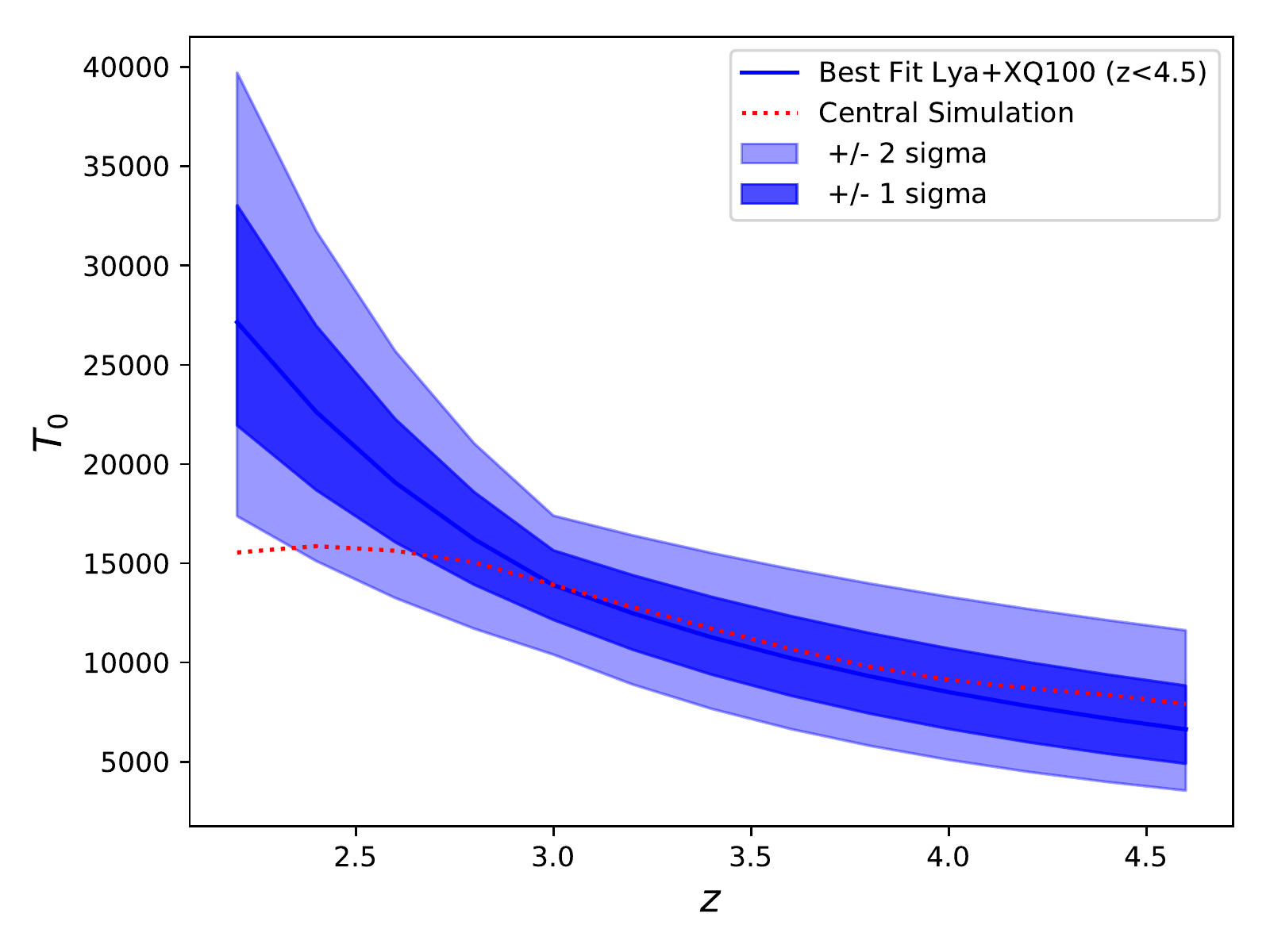,width = .49\textwidth}
\epsfig{figure= 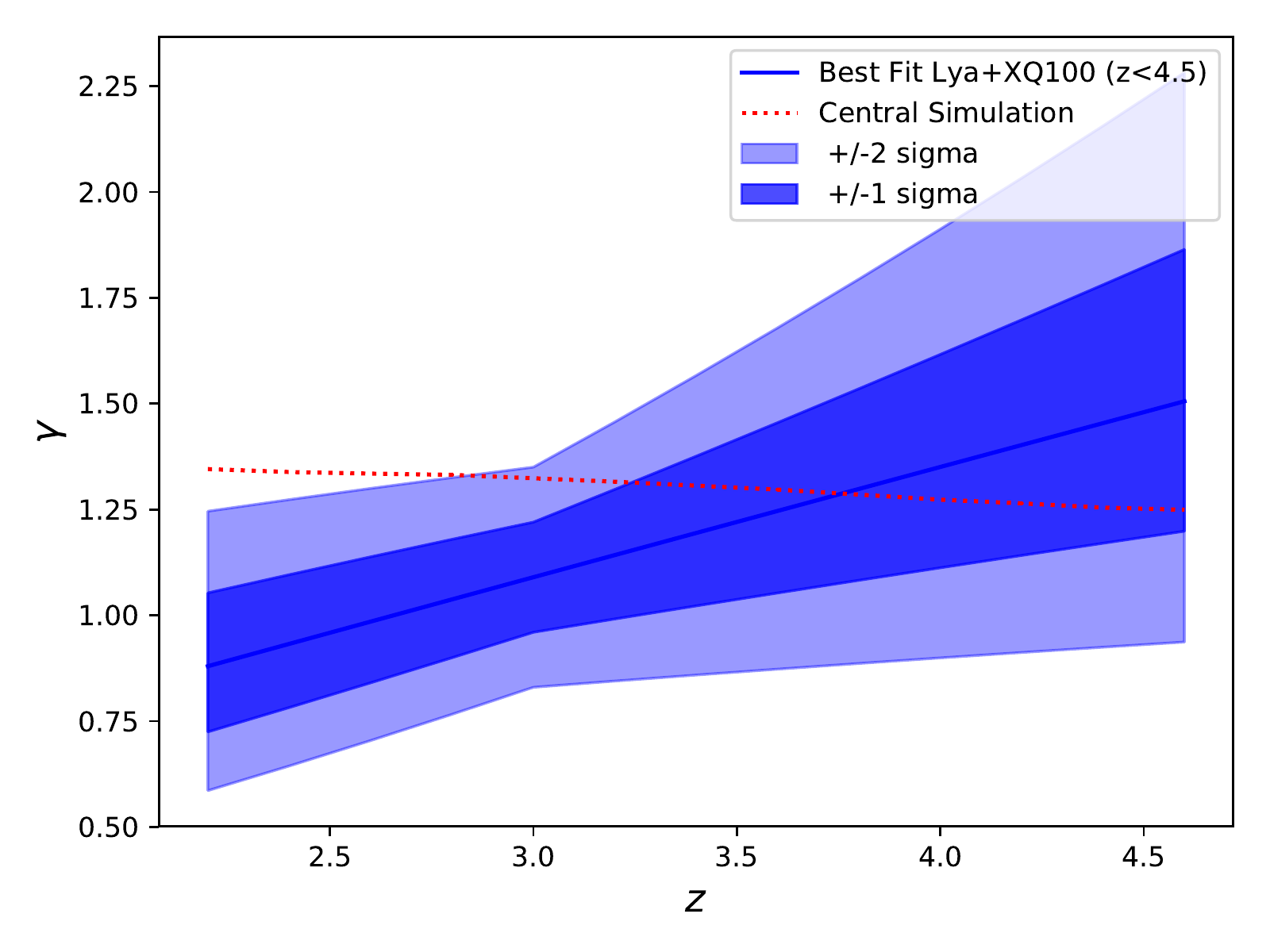,width = .49\textwidth}\\
\epsfig{figure= 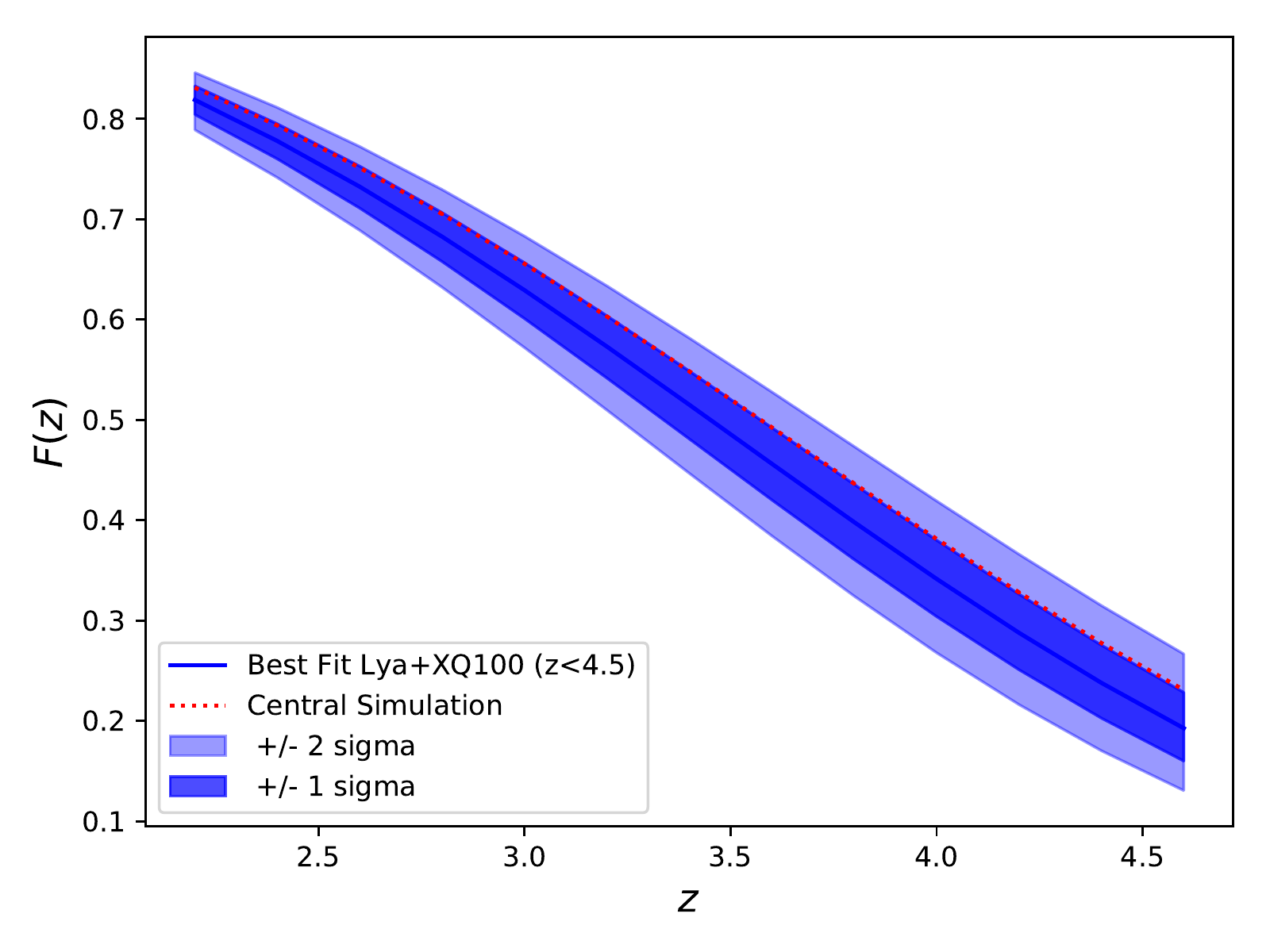,width = .49\textwidth}
\caption{Thermal history for the $m_X$ fit. Same color code as figure~\ref{fig:thermal_mnu}.}
\label{fig:thermal_wdm}
\end{center}
\end{figure}

We model the IGM thermal history  through a set of parameters given in table~\ref{tab:astroparam}. The recovered range of the IGM temperature $T(z)$, density dependence $\gamma(z)$ and mean flux $\overline{F}(z)$ are shown in figure~\ref{fig:thermal_mnu} for the fits on the sum of the neutrino masses $\sum m_\nu$, and in figure~\ref{fig:thermal_wdm} for the fit on the mass of a warm dark matter thermal relic. Current measurements of these parameters exhibit a wide spread.  Values of $T_0$  vary among authors by over a factor 2 (typically between about $10\,000$K and $20\,000$K for redshifts between 2 and 3, decreasing with redshift to  around $8\,000$K at $z=4$), and $\gamma$ can be found between  1.0 and 1.8~\cite{Hiss2018,Lidz2014,Walther2019}.  The central simulation of our simulation grid corresponds to the thermal model of \cite{Becker2011}.
For the purpose of this study, thermal parameters are taken as nuisance and we do not try to recover precise values. Instead, we allow for a lot of freedom in their range and modeling.

In all fits, the recovered thermal history overlaps with the typical range allowed by observations. Warm dark matter fits are in perfect agreement with the central simulation on all three parameters. Neutrino mass fits tend to have a cold IGM at high redshift,  although the 2 $\sigma$ bound is within acceptable limits. Planck data do not constrain the IGM thermal history. However, we note that adding them reduces the uncertainty on the thermal parameters through their correlation with cosmological parameters. The main impact is for $\gamma$, where its anti-correlation with $\Omega_M$ and $n_s$, combined with the light tension between \lya and CMB data on these parameters, pushes it to low values in the \lya+P18 neutrino mass fit. In contrast, \lya data alone yield perfectly standard values of $\gamma$, $T_0$ and $\overline{F}(z)$. 

To further check the robustness of our results, we tested a different, less flexible, thermal model. Instead of the one described by the parameters of table~\ref{tab:astroparam} that leave the slopes of the redshift-dependence free, we fixed the shape of the thermal history  to that of~\cite{Becker2011} used in the central simulation, and we only allowed changes in amplitude and density dependence of the UV heating rates (AMPL and GRAD parameters of {\sc Gadget}), effectively changing $T_0$ and $\gamma$ at all redshifts.
The limits obtained in this case are summarized in table~\ref{tab:new_thermo}. We note that the recovered thermal history is in excellent agreement with that of the central simulation in all cases. While considerably reduced, the trends noticed above when we included Planck data are still true (in particular a slightly lower value of $\gamma$). The bound on $\sum m_\nu$ is slightly tightened when imposing the more constrained thermal model. The looser fit of the paper therefore yields conservative bounds. The warm dark matter results are only mildly changed.

\begin{table}[htbp]
\begin{center}
\begin{tabular}{lccc}
\hline\hline
Configuration &  Constraint {\scriptsize(95\% CL)} & $T_0(z=3)$ & $\gamma(z=3)$\\
 \hline \\[-10pt]

\lya   &  $\sum m_\nu < 0.39$~eV & 7900K & 1.1\\ [2pt]
\lya + P18 &  $\sum m_\nu < 0.07$~eV & 15700 & 1.0\\[2pt]
\lya ($z\!<\!4.5$) + XQ100 & $m_X>4.7$~keV & 13900 & 1.1\\[2pt]
\hline
\end{tabular}
\caption{\label{tab:new_thermo} Constraints on the sum of the neutrino masses and on the mass of WDM in the form of thermal relics assuming a thermal history following a similar shape as that of~\cite{Becker2011}. The last two columns give the equivalent $T0$ and $\gamma$ at redshift 3 for the best-fit model.}
\end{center}
\end{table}

\section{Correlation coefficients}
\label{sec:appendixB}
We provide the correlation coefficients between  fit parameters in the case of the $\sum m_\nu$ fit to \lya data alone in table~\ref{tab:corcoef}, and in the case of the $1\,{\rm keV}/m_X$ fit to \lya $(z\!<\!4.5)$ + XQ100 data in  table~\ref{tab:corcoefWDM}. For the sake of clarity, we restrict the table to matrix rows and columns that exhibit at least one coefficient with absolute value exceeding 25\%, and we only show coefficients larger than 20\% in absolute value.

\begin{table}[htbp]
\begin{center}
\tiny
\begin{tabular}{r|rrrrrrrrrrrrr}
& \multirow{2}{*}{$A^\tau$} & \multirow{2}{*}{$\eta^\tau$} & \multirow{2}{*}{$\sigma_8$} & \multirow{2}{*}{$n_s$} & \multirow{2}{*}{$T_0$} & \multirow{2}{*}{$\gamma$} & \multirow{2}{*}{$\Omega_m$} & \multirow{2}{*}{$H_0$} & $\eta^T$ & $\eta^T$ & \multirow{2}{*}{$\eta^\gamma$} & \multirow{2}{*}{$A^{\rm splice}$} & \multirow{2}{*}{$A^{\rm AGN}$} \\
&&&&&&&&&$z\!<\!3$ & $z\!>\!3$ \\
&	 0&	 1&	 2&	 3&	 4&	 5&	 6&	 7&	 8&	 9&	 10&	 11&	 12\\
 \hline \\[-5pt]
0&	1.\\
1&	-0.46&	1.\\
2&	 -- &	 -- &	1.\\
3&	-0.48&	 -- &	 -- &	1.\\
4&	 -- &	-0.27&	-0.35&	0.38&	1.\\
5&	0.25&	0.39&	 -- &	-0.30&	-0.60&	1.\\
6&	 -- &	 -- &	0.39&	-0.37&	-0.23&	-0.21&	1.\\
7&	 -- &	 -- &	 -- &	-0.30&	 -- &	 -- &	 -- &	1.\\
8&	-0.34&	 -- &	-0.43&	0.43&	0.62&	 -- &	-0.34&	 -- &	1.\\
9&	 -- &	 -- &	 -- &	 -- &	 -- &	0.54&	 -- &	 -- &	0.25&	1.\\
10&	0.62&	-0.58&	 -- &	 -- &	0.24&	-0.51&	 -- &	 -- &	-0.41&	-0.21&	1.\\
11&	 -- &	0.64&	-0.25&	 -- &	 -- &	0.27&	-0.29&	 -- &	 -- &	 -- &	 -- &	1.\\
12&	 -- &	 -- &	 -- &	-0.39&	 -- &	 -- &	 -- &	 -- &	-0.28&	-0.22&	 -- &	 -- &	1.\\
\end{tabular}
\normalsize

\caption{\label{tab:corcoef} Correlation coefficients for the $\sum m_\nu$ fit on \lya. Only parameters with at least one correlation coefficient exceeding 25\% in absolute value are shown, and we display all coefficients smaller than 0.20 in absolute value as \tquote{ -- }. }
\end{center}
\end{table}

\begin{table}[htbp]
\begin{center}
\tiny
\begin{tabular}{r|rrrrrrrrrrrrrrr}
& \multirow{2}{*}{$A^\tau$} & \multirow{2}{*}{$\eta^\tau$} & \multirow{2}{*}{$n_s$} & \multirow{2}{*}{$T_0$} & \multirow{2}{*}{$\gamma$} & \multirow{2}{*}{$\Omega_m$} & \multirow{2}{*}{$H_0$} & \multirow{2}{*}{$\frac{1}{m_X}$} &$\eta^T$ & $\eta^T$ & \multirow{2}{*}{$\eta^\gamma$} & \multirow{2}{*}{$A^{\rm splice}$} & \multirow{2}{*}{$A^{\rm AGN}$}  & \multirow{2}{*}{$A^{\rm SN}$} & \multirow{2}{*}{$z_{\rm reio}$}\\
&&&&&&&&&$z\!<\!3$ & $z\!>\!3$ \\
&	 $0$&	 $1$&	 $2$&	 $3$&	 $4$&	 $5$&	 $6$&	 $7$&	 $8$&	 $9$&	 ${10}$&	 ${11}$&	 ${12}$&	 ${13}$&	 ${14}$\\
 \hline \\[-5pt]
0&	1.	\\
1&	-0.75	&1.	\\
2&	-0.45	&0.42	&1.	\\
3&	 -- 	& -- 	& -- 	&1.	\\
4&	 -- 	&0.24	& -- 	&-0.84	&1.	\\
5&	 -- 	& -- 	&-0.41	& -- 	& -- 	&1.	\\
6&	 -- 	& -- 	&-0.27	& -- 	& -- 	& -- 	&1.	\\
7&	0.27	&-0.32	&-0.21	& -- 	& -- 	& -- 	& -- 	&1.	\\
8&	 -- 	&-0.35	& -- 	& -- 	& -- 	& -- 	& -- 	& -- 	&1.	\\
9&	 -- 	& -- 	& -- 	&-0.82	&0.72	& -- 	& -- 	& -- 	& -- &1.	\\
10&	0.37	& -- 	& -- 	& -- 	&-0.40	& -- 	& -- 	& -- 	&-0.49	&-0.28	&1.	\\
11&	 -- 	&0.46	&0.33	& -- 	&-0.20	& -- 	& -- 	& -- 	&-0.51	&-0.32	&0.62	&1.	\\
12&	 -- 	& -- 	&-0.37	& -- 	& -- 	& -- 	& -- 	& -- 	&-0.38	& -- 	&0.21	& -- 	&1.	\\
13&	 -- 	& -- 	&0.40	& -- 	& -- 	& -- 	& -- 	& -- 	& -- &-0.31	& -- 	& -- 	& -- 	& -- 	\\
\end{tabular}
\normalsize

\caption{\label{tab:corcoefWDM} As table~\ref{tab:corcoef} for the WDM fit on \lya $(z\!<\!4.5)$ + XQ100. }
\end{center}
\end{table}

In the $\sum m_\nu$ fit, the cosmological parameters show correlation with some of the astrophysical or nuisance parameters, but never at a level exceeding 50\%. The largest such correlations are between $n_s$ and $A^\tau$ (-48\%), between $\sigma_8$ and $\eta^T(z<3)$ (-43\%), and between $n_s$ and $\eta^T(z<3)$ (43\%). All other correlations with cosmological parameters are below 40\%. In contrast, astrophysical and nuisance parameters show larger degeneracies, often in excess of 50\% correlations. This is why, in particular, we do not have any claims on thermal history but instead consider parameters that describe it as nuisance, which we marginalize over (in the Bayesian approach) or leave free in the fit (in the frequentist approach).  

In the WDM fit, $m_X$ is only slightly correlated to the mean flux (27\% with $A^\tau$ and $-32\%$ with $\eta^\tau$) and to $n_s$ ($-21\%$). We find no significant correlation of $m_X$ with any of the parameters describing the thermal history.

\bibliographystyle{JHEP}
\bibliography{bib}{}
\end{document}